\documentclass[usenatbib]{mnras}

\usepackage[T1]{fontenc}

\DeclareRobustCommand{\VAN}[3]{#2}
\let\VANthebibliography\thebibliography
\def\thebibliography{\DeclareRobustCommand{\VAN}[3]{##3}\VANthebibliography}

\usepackage{hyperref}
\usepackage{graphicx}	
\usepackage{amsmath}	
\usepackage{amssymb}	

\usepackage{xcolor}  

\newcommand{\msun}{$M_{\odot}$}
\newcommand{\msolar}{M_{\odot}}

\newcommand{\mtot}{m_{\mathrm{tot,min}}}
\newcommand{\mgas}{m_{\mathrm{gas,min}}}
\newcommand{\zgas}{Z_{\mathrm{max}}}
\newcommand{\zsun}{$Z_{\odot}$}
\newcommand{\hinv}{$h^{-1}$}

\title[Building Black Hole Seeding Models Using IllustrisTNG]{Building Semi-Analytic Black Hole Seeding Models Using IllustrisTNG Host Galaxies}

\author[Analis Eolyn Evans et.~al.]{
Analis Eolyn Evans$^{1}$\thanks{analislawrence@ufl.edu},  Laura Blecha $^{1}$,
Aklant Kumar Bhowmick $^{1}$
\\
$^{1}$Department of Physics, University of Florida, Gainesville, FL 32611, USA\\
}

\pubyear{2023}

\begin{document}
\label{firstpage}
\pagerange{\pageref{firstpage}--\pageref{lastpage}}
\maketitle

\begin{abstract}

Because early black holes (BHs) grew to $\sim10^{9}$ \msun\ in less than 1 Gyr of cosmic time, BH seeding models face stringent constraints. 
To efficiently constrain the parameter space of possible seeding criteria, we combine the advantages of the cosmological IllustrisTNG~(TNG) simulations with the flexibility of semi-analytic modeling. We identify TNG 
galaxies as BH seeding sites based on various criteria including a minimum gas mass of $10^7$-$10^9$ \msun, total host mass of $10^{8.5}$-$10^{10.5}$ \msun, and a maximum gas metallicity of 0.01 - 0.1 \zsun. Each potential host is assigned a BH seed with a probability of 0.01 - 1; these BHs are then traced through the TNG galaxy merger tree. This approach improves upon the predictive power of the simple TNG BH seeding prescription, especially in the low-mass regime at high redshift, and it is readily adaptable to other cosmological simulations. Most of our seed models predict $z\lesssim4$ BH mass densities that are consistent with empirical data as well as the TNG BHs. However, high-redshift BH number densities can differ by factors of $\sim$ 10 - 100 between models. 
In most models, $\lesssim10^5$ \msun\ BHs substantially outnumber heavier BHs at high redshifts. Mergers between such BHs are prime targets for gravitational-wave detection with LISA. 
The $z=0$ BH mass densities in most models agree well with observations, but our strictest seeding criteria fail at high redshift. Our findings strongly motivate the need for better empirical constraints on high-$z$ BHs, and they underscore the significance of recent AGN discoveries with JWST. 
\end{abstract}


\section{Introduction}

 Observations of luminous active galactic nuclei (AGN) at $z \sim$ 6-11 \citep{2006ARA&A..44..415F,2009AJ....138..305J,2011Natur.474..616M,2018ApJ...856L..25B, 2021ApJ...907L...1W,2022Natur.604..261F, 2023arXiv230512492M, 2023arXiv230308918L,2023ApJ...942L..17O} 
 indicate that BHs assembled in less than $\sim 0.5$-1 Gyr of cosmological time after the Big Bang. 
This poses significant challenges for current BH seeding and growth models.
For example, the earliest Population III (Pop III) stars were massive and essentially metal-free, and they should have therefore created massive BH remnants. Population III stars form in the gravitational potential well of dark matter mini-halos that collapse at $z\sim 20$ due to high matter density fluctuations; these are expected to form $\approx10^{2}-10^{3}$ \msun\ BH seeds \citep{2003ApJ...582..559V}. However, such seeds would require sustained periods of super-Eddington accretion to reach the supermassive regime by the epoch of the earliest quasars. 

A possible solution to these tight constraints on BH growth timescales is that seeds form at higher initial masses. One promising scenario is that of direct collapse BHs (DCBHs),  seeded when a massive, metal-free gas cloud collapses directly into a BH or supermassive star (SMS) with a mass of $\approx10^{5}$ \msun\ \citep[e.g.,][]{1984ARA&A..22..471R,2003ApJ...596...34B,2006MNRAS.370..289B,2020ARA&A..58...27I}. This monolithic collapse must be aided by dissociation of molecular hydrogen through UV radiation in the Lyman-Werner band, which could be provided by nearby star-forming regions. Lyman-Werner radiation prevents the fragmentation that would normally happen in low-temperature gas near the cosmological Jeans mass \citep{2003ApJ...596...34B}. Alternately, dynamical heating by mergers or turbulent cold flows may suffice to suppress fragmentation and allow DCBH formation \citep[e.g.,][]{2010Natur.466.1082M,2019Natur.566...85W,2022Natur.607...48L,2023MNRAS.518.2076Z}. Magnetic fields have been proposed as another means to catalyze and trigger the formation and early growth of massive BH seeds, by suppressing fragmentation and star formation and boosting the accretion flow to newly formed DCBH seeds  \citep{2023arXiv230519081B}.
Any mergers between heavy DCBH seeds would create an additional avenue for growth, and these would be prime candidates for gravitational-wave (GW) detection with the Laser Interferometer Space Antenna (LISA) \citep{2017arXiv170200786A,2023LRR....26....2A}.  
	
 Additionally, SMBH seeds can naturally be formed via successive mergers of Pop III stellar remnants, which could form intermediate-mass BHs (IMBHs) of $\approx10^{3}-10^{5}$ \msun\ \citep{1984ApJ...280..825B,2001ApJ...551L..27M,2011ApJ...740L..42D,2022MNRAS.511.2631A}. In the nearby Universe, many BHs co-exist with dense, massive regions of stars near galactic centers known as nuclear star clusters (NSCs), kindling the idea that NSCs may have come before the SMBH \citep[][and references therein]{2020A&ARv..28....4N,2022MNRAS.511.2631A}. The formation pathways of IMBHs within NSCs depend on the their mass, density, and spin \citep{2002MNRAS.330..232C, 2020ARA&A..58..257G, 2020MNRAS.498.4591F}. 
 Merger events in nuclear star clusters are potential GW sources for 
 LISA as well as for ground-based GW detectors such as the LIGO-Virgo-KAGRA (LVK) collaboration \citep{2016PhRvD..93l2003A,2022arXiv220311781J}. Alternatively, a $10^3-10^5$~\msun\ BH seed \citep{2015ApJ...810...51M} could be formed through a supermassive Pop III stellar remnant \cite[e.g.,][]{2004ARA&A..42...79B,2003ApJ...591..288H,2015MNRAS.448.1835T} but this scenario is also uncertain and would likely still involve super-Eddington growth.

Not much is understood about early-Universe BH-galaxy co-evolution due to
observational limitations. For example, selection biases make it unclear how 
the BH mass - stellar mass relation may evolve above $z\sim2$ \citep{2003ApJ...583..124S,2009ApJ...706L.215J,2020ApJ...889...32S}. One key observational bias is that it is difficult to observe faint quasars especially at high redshift, which could reveal more about the entire BH population than bright quasars \citep{2022MNRAS.511.3751H}. Accordingly, AGN observations show increasing uncertainties in the BH mass function (BHMF) up to $z\sim6$  \citep{2008MNRAS.388.1011M,2009ApJ...690...20S,2010ApJ...725..388C}. The faint, early-Universe quasars can help explain the stepping stones for assembly of the massive, most luminous ones through constraining early demographics. 
JWST can detect rest-frame UV and optical light from faint quasars that has previously been inaccessible for high-redshift quasars \cite[e.g.,][]{2012ApJ...756..150D,2020ApJ...900...21M}. There have already been numerous high-redshift BH candidates identified in JWST data, including low-mass candidates \citep{2023ApJ...942L..17O,2023arXiv230512492M, 2023arXiv230308918L,2023arXiv230200012K,2023arXiv230206647U,2023arXiv230311946H,2023arXiv230605448M, 2023arXiv230607320L, 2023MNRAS.525.1353J, 2023arXiv230801230M}. 
This trove of early discoveries is a promising indication that JWST will continue to reveal a great deal about early BH-galaxy co-evolution and the epoch of reionization.

As noted above, BH assembly mechanisms are also of great interest owing to the potential for GW detections of BH mergers with LISA and LVK, as well as next-generation ground-based GW detectors. LISA will be revolutionary for our understanding of BH assembly in a regime where electromagnetic (EM) constraints are sparse or non-existent, with the capacity to detect BH mergers in the mass range $\sim10^{4}\relbar10^{7}\msolar / (1+z)$
out to $z\sim20$ \citep{2004SPIE.5500..183V,2006PhRvD..74l2001L,2007PhRvD..75h9902L,2017arXiv170200786A,2023LRR....26....2A}. 

Pulsar Timing Array (PTA) experiments are sensitive to GWs in the $\lesssim$ nanoHertz - microHertz range, corresponding to $\sim 10^9$\msun\ BH binaries. Recently, PTAs around the globe presented strong evidence for a stochastic GW background that is consistent with the expected signal from a cosmological population of BH binaries \citep{2023ApJ...951L...8A,2023arXiv230616214A,2023ApJ...951L...6R,2023RAA....23g5024X}. Future PTA data will constrain the spectral shape of this background and the (an)isotropy of its origin on the sky, both of which will provide key insight into SMBH binary evolution.

Improved predictions of GW and EM signatures from different BH assembly channels are needed to interpret data from the upcoming observations described above. Many theoretical studies of BH formation and growth rely on semi-analytic models (SAMs), which have the unique ability to probe a wide range of seeding scenarios with little computational expense~\citep[e.g.,][]{2007MNRAS.377.1711S,Volonteri_2009, 2012MNRAS.423.2533B,2018MNRAS.476..407V, 2018MNRAS.481.3278R, 2019MNRAS.486.2336D,2021MNRAS.506..613S}. Most of these SAMs have thus far relied on tracking BH seeding and growth over halo merger trees constructed using analytic formulations such as the Press-Schechter~\citep{1974ApJ...187..425P} or dark-matter-only cosmological simulations. However, by construction, SAMs cannot trace the detailed hydrodynamics of the gas or the internal structure of galaxies. This poses as a significant limitation on modeling BH seed formation, which crucially relies on the local gas conditions within halos. 

Alternatively, BH evolution can also be modeled in cosmological hydrodynamics simulations, which~(unlike SAMs) do solve the gas hydrodynamics along with sub-grid prescriptions for BH seeding, accretion, and feedback. Numerous large-volume cosmological simulations including Illustris, IllustrisTNG (hereafter TNG), SIMBA, EAGLE, and Horizon-AGN have been shown to produce results consistent with many observed properties of galaxy and BH populations, including the BH-bulge relation \citep{2014Natur.509..177V, 2014MNRAS.444.1453D, 2015MNRAS.446..521S, 2017MNRAS.465.3291W, 2018MNRAS.475..648P, 2019MNRAS.486.2827D}. However, these simulations have a major drawback compared to SAMs, in that their huge computational expense prohibits exploring large parameter spaces. Further, most of these simulations still cannot directly resolve low-mass BH seeds.

Due to these challenges, most large-volume cosmological simulations adopt very simplistic seed models. For example, many simulations seed $\sim10^5-10^6~M_{\odot}$ BHs in halos above a fixed mass threshold of $\sim10^9-10^{10}~M_{\odot}/h$~\citep{2014Natur.509..177V,2015MNRAS.450.1349K,2015MNRAS.446..521S,2016MNRAS.455.2778F,2019ComAC...6....2N}. While these
simple prescriptions reproduce local BH populations reasonably well, 
their predictive ability at high redshift is limited, and they cannot distinguish between different BH seeding channels. Several simulations have also seeded BHs based on local gas properties in cosmological simulations~\citep[e.g.,][]{2014MNRAS.442.2751T,2017MNRAS.470.1121T,2019MNRAS.483.4640W}. In particular, they create BH seeds from gas cells that exceed a critical density threshold while remaining metal-poor. These prescriptions are much more representative of theoretical seeding channels such as Pop III, NSC and DCBH seeds, all of which are expected to form exclusively in regions of dense and metal-poor gas. However, at coarse gas mass resolutions, the poor resolution convergence of gas properties could impact of resolution convergence of the final BH populations. 

At very high gas mass resolutions~($\sim10^3-10^4~M_{\odot}$) typical of zoom simulations, gas-based seed models do start producing reasonably well converged BH populations~\citep{2021MNRAS.507.2012B}. Zoom simulations are also relatively computationally inexpensive, which has allowed several recent works to explore a wide range of gas-based seed models~\citep{2021MNRAS.507.2012B,2022MNRAS.510..177B,2022MNRAS.516..138B}. However, since zoom simulations typically focus on a small biased region of the universe, they cannot be readily compared to observations.

In this work, we adopt a new approach that harnesses the strengths of conventional SAMs and full hydrodynamics simulations, while mitigating the limitations inherent in each approach. We develop novel, SAM-based BH seed models that can trace BH evolution across merger trees within any existing cosmological hydrodynamical simulation. By doing this, our seeding prescriptions can be informed by the detailed gas properties of halos on the merger trees, which are inaccessible in conventional SAMs. A similar approach was taken by \cite{DeGraf_2019} wherein BH growth histories within Illustris were reconstructed for subsets of the simulated BH population. These subsets were selected by introducing additional seeding criteria beyond the default seed model used by Illustris, such as spin and metallicity based seeding. They found that the total BH merger rate can be substantially impacted by the introduction of these seeding criteria. In contrast to \cite{DeGraf_2019}, our models place new seed BHs in the subhalo merger trees that are completely independent of the BHs that formed during the actual run of the parent simulation. This enables us to study a wide variety of BH formation models, including criteria that are more lenient than those used on-the-fly during the simulation run.  

For our parent simulation, we use the highest-resolution run of the TNG suite, TNG50-1. In Appendix~\ref{sec:res_converge}, we use the lower-resolution versions of TNG50 for convergence tests. Unless otherwise specified, ``TNG50'' refers to the highest-resolution TNG50-1 simulation in the remainder of the paper. With a gas mass resolution of $8\times10^4~M_{\odot}$, TNG50 offers a resolution comparable to zoom simulations over a reasonably large volume of $(50~\mathrm{Mpc})^3$ \citep{Nelson_2019}. This allows us to use well resolved gas properties to design and explore a large ensemble of new seed models motivated by proposed theoretical seeding channels, which produce BH populations that can be compared to observations. The model assumptions include allowing a maximum of one BH per massive, low-metallicity, galaxy or galaxy group with a sufficient gas reservoir. A key advantage is that the stellar, gas, and host properties that inform the seed models are directly attainable from the simulation. This avoids the use of an empirical framework to derive the baryonic properties, which is commonly used in most SAMs. 

This paper is organized as follows. Section \ref{sec:methods} summarizes key features of the IllustrisTNG simulations, describes our methodology for constructing a TNG-based SAM for BH seeding and growth, and details the parameter space explored in this work. Section \ref{sec:results} present our results, including an analysis of the properties of high-$z$ TNG halos (Section \ref{sec:massmet_relations}), a verification that our SAM can successfully reproduce the TNG BH population (Section \ref{tng_vs_model}), and a detailed analysis of the BH populations produced by our SAM, including BH number and mass density evolution and local BHMFs (Sections \ref{ssec:fiducial} \& \ref{ssec:bhmf}). We summarize and conclude in Section \ref{sec:conclude}. Throughout this paper, we assume the same cosmology as the TNG simulation suite (as specified below).

\section{Methods}
\label{sec:methods}
\subsection{IllustrisTNG simulations}

The TNG simulation project is a  
cosmological magnetohydrodynamical simulation suite 
\citep{2018MNRAS.480.5113M,2018MNRAS.475..648P,2018MNRAS.475..676S,2018MNRAS.477.1206N}. The initial cosmological conditions are
$\Omega_{\Lambda,0}=0.6911$, $\Omega_{m,0}=0.3089$, $\Omega_{b,0}=0.0486$, $\sigma_{8}=0.8159$, $n_{s}=0.9667$, and $h=0.6774$,  taken from Planck collaboration observations of the cosmic microwave background \citep{2016}. 
These simulations were carried out with the quasi-Lagrangian AREPO code~\citep{2010ARA&A..48..391S,2011MNRAS.418.1392P,2013MNRAS.432..176P,2020ApJS..248...32W} in which gravitational equations are coupled with magnetohydrodynamics~(MHD) equations. The gravity is solved using a tree-particle-mesh N-body algorithm, and the MHD is solved using
an adaptive unstructured mesh that is constructed by performing a Voronoi tesselation of the simulation volume.

AREPO implements sub-grid modeling for a variety of physical processes that cannot be directly resolved in current cosmological simulations. These include gas cooling, star formation and evolution, chemical enrichment and feedback. Star formation happens within gas above a critical threshold density of $0.1~\mathrm{cm}^{-3}$ \citep{2003MNRAS.341.1253H}. Stellar evolution assumes an initial mass function from \cite{2003PASP..115..763C}, which leads to their metal enrichment. The stellar feedback includes energy released from AGB stars and supernovae, and it is primarily responsible for depositing metals on to the surrounding gas. Further details about the implementation of these processes are described in \cite{2018MNRAS.473.4077P}. 
The BH-related sub-grid physics models will be discussed in more detail below and in Sections~\ref{sec:BH_form_evol} and ~\ref{sec:merger_trees}.

Subhalo and halo catalogs are saved for each snapshot with with a wide range of quantities including gas-phase metallicities, star-formation rates, stellar, BH, and total host masses, velocity dispersion, and the number of BHs per subhalo or halo. The Friends-of-Friends (FoF) algorithm \citep{1982ApJ...259..449P,1982ApJ...257..423H,2005ApJ...630..759M} groups DM particles together if they are within $0.2$ times the mean separation \citep{2015MNRAS.452.2247V}. Therefore, the halos can be generally identified as groups of galaxies. The subhalo catalog is computed using \texttt{SUBFIND} \citep{2001MNRAS.328..726S}; subhalos can generally be identified as galaxies in the simulation.
For a negligible number of catalog objects near the resolution limit, the algorithm cannot distinguish galaxies versus spurious clumps; these are excluded from our analysis based on their tendency to have very low masses.

TNG has overall produced good agreement for BH and galaxy properties, including, but not limited to, BH scaling relations \citep{2020ApJ...895..102L}, correlations between SMBH mass and X-ray temperature of the hot gaseous halos pervading host galaxies, the underlying SMBH-halo mass relation \citep{2021MNRAS.501.2210T}, the BH-stellar bulge mass relation \citep{2017MNRAS.465.3291W,2021MNRAS.503.1940H}, and anisotropic black hole feedback causing quiescent satellites to be found less frequently along the minor axis of their central galaxies \citep{2021Natur.594..187M}.
Our primary simulation TNG50-1 has a (50 Mpc)$^3$ box that includes $2160^{3}$ gas cells  
\citep{Nelson_2019}.

\subsubsection{BH formation and evolution}
\label{sec:BH_form_evol}

Seeding, growth, and feedback are all important processes in BH evolution. In the TNG simulation, BHs of seed mass $8\times10^{5}$ \hinv \msun\ are placed in halos with dark matter halos exceeding a total mass threshold of $5\times10^{10}$ \hinv \msun\ \citep{2017MNRAS.465.3291W}. More specifically, the densest gas particle of a halo is converted to a BH particle if the halo does not already contain a BH. 

BH growth is modeled by Eddington-limited Bondi accretion (and can also be facilitated through mergers):
\begin{eqnarray}
\dot{M}_{\rm{Edd}}=\frac{4\pi G M_{\rm{BH}} m_p c}{\epsilon_r \sigma_T},\\
\dot{M}_{\rm{Bondi}}=\frac{4 \pi G^2 M_{\rm{BH}}^2 \rho}{c_s^3},\\
\dot{M}_{\rm{BH}}=\rm{min}(\dot{M}_{\rm{Bondi}}, \dot{M}_{\rm{Edd}}),
\end{eqnarray}

where $M_{\rm BH}$ is the BH mass, $\epsilon_r$ is the radiative efficiency (set to $0.2$ in TNG), $\sigma_T$ is the Thomson scattering cross-section, $m_p$ is the proton mass, and $\rho$ \& $c_s$ are the gas density and sound speed, respectively, in cells neighboring the BH. The feedback model for BHs in TNG  assumes thermal or kinetic energy feedback modes from the AGN. The kinetic mode is comparably more efficient and is the dominant means for SMBH growth for BHs above $\approx10^{8}$ \msun\ at low accretion rates relative to the Eddington limit \citep{2017MNRAS.465.3291W}. The thermal mode of AGN feedback is associated with high accretion rates and jets, where along with mergers, it is responsible for the star-formation quenching of massive galaxies \citep{2017MNRAS.465.3291W}.
	
\subsubsection{Merger Trees}
\label{sec:merger_trees}

The Sublink merger trees
\citep{2015MNRAS.449...49R} include a descendants tree branch with galaxy identifiers that allow merger tracking.  
TNG descendant selection is performed by first identifying subhalo descendant candidates, scoring them with a merit function based on the particle's binding energy rank, and deeming the descendant as the one with the highest score \citep{2015MNRAS.449...49R}.
Following TNG's critical descendant links in our reconstructed TNG merger trees, starting from points at which galaxies meet the model seeding criteria, we are able to follow these populations of galaxies and their BHs, each with its own unique merger history.

\subsection{Simulation Analysis: Semi-Analytic Black Hole Seeding Model}

\subsubsection{Identifying BH seeding sites}
\label{sec:seedingsites}
For the novel, hybrid SAMs, we apply host criteria to identify BH seeding sites within TNG in a post-processing approach.
Gas mass and metallicity properties in TNG halos are examined, since all gas-based BH seeding models require low metallicity as well as a large enough gas reservoir to form seeds.

\begin{figure*}
    \centering
    \includegraphics[width=0.95\columnwidth]{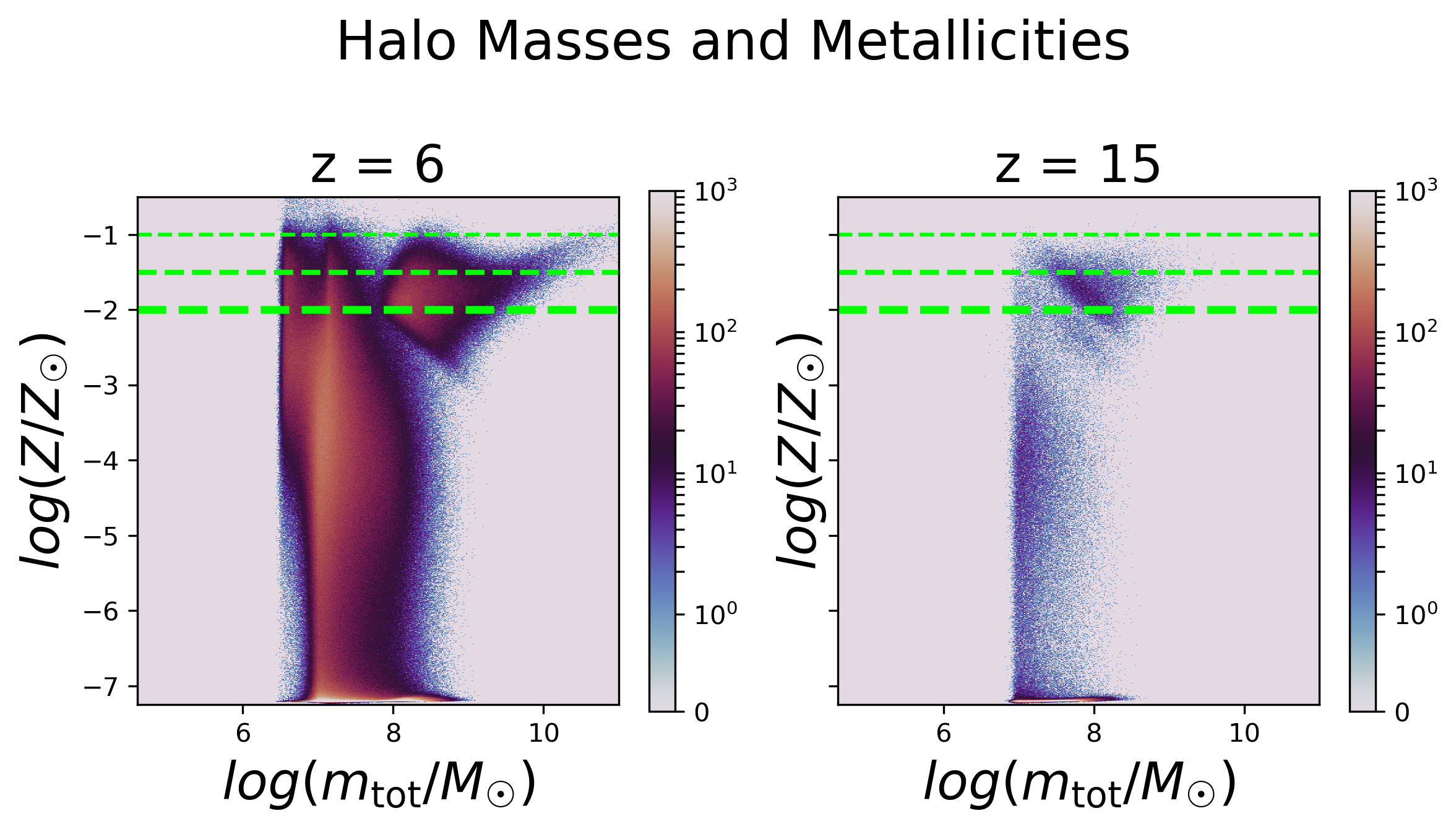}
    \includegraphics[width= 0.95\columnwidth]{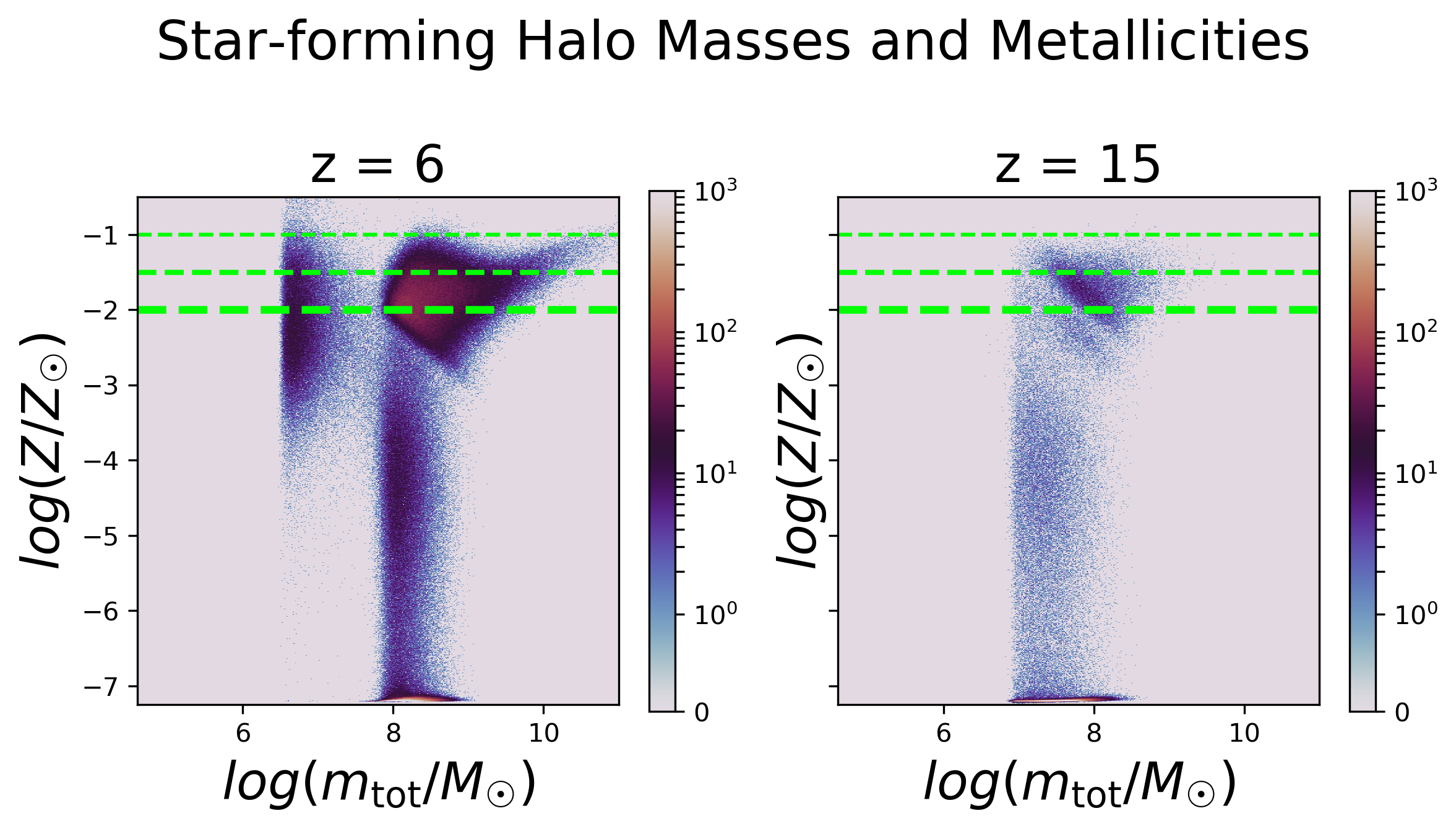}
    \includegraphics[width=0.95\columnwidth]{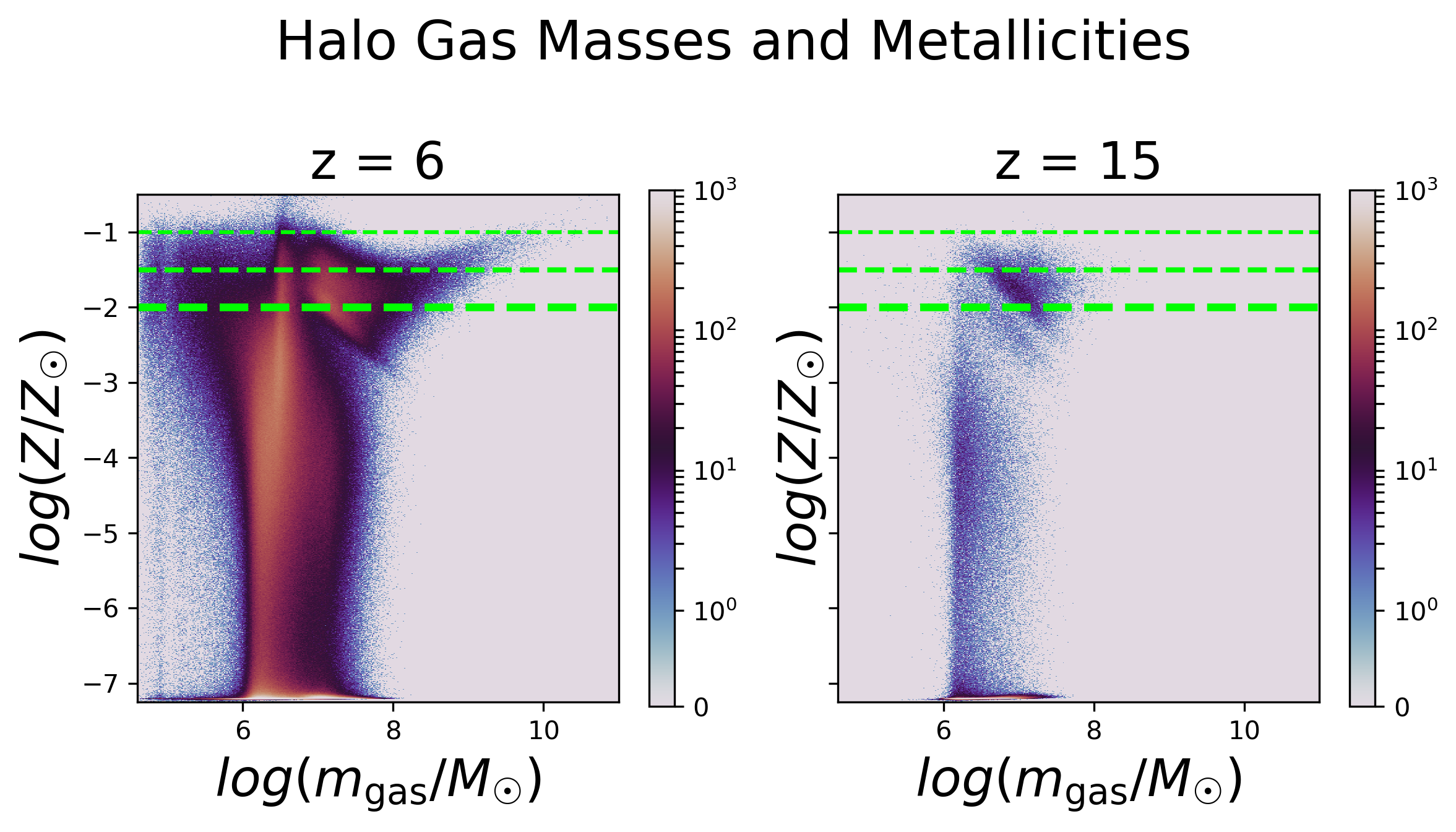}
    \includegraphics[width=0.95\columnwidth]{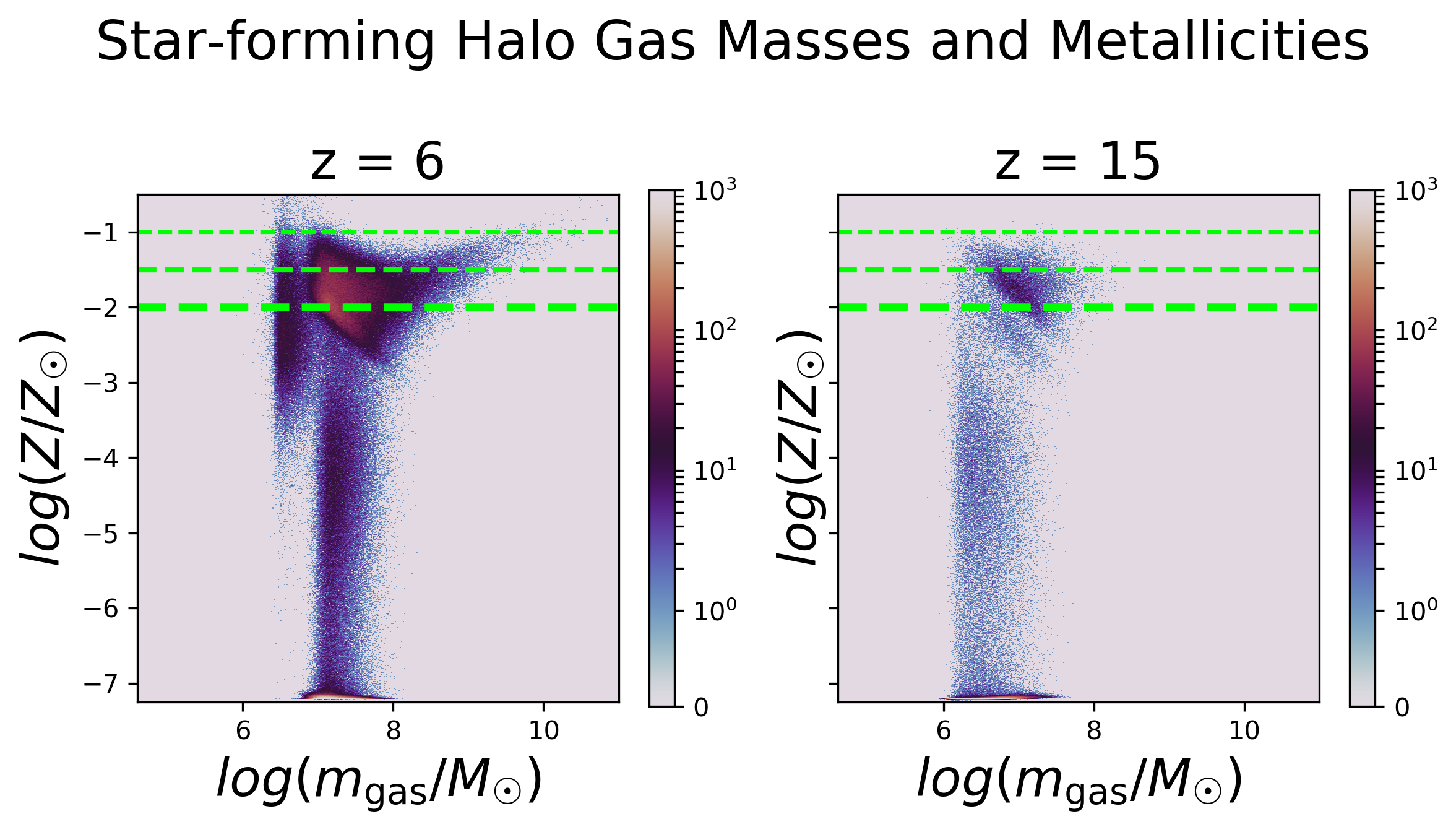}
   
    \caption{We use 
    2D histograms of gas metallicity versus gas mass and versus total mass to illustrate the properties of high-redshift halos; these distributions motivate our choices of seeding criteria. The maximal gas metallicity values used in our models are 
    $\zgas/$\zsun~$=10^{-1}$, $10^{-1.5}$, or $10^{-2}$ (shown by the thinnest to thickest dashed green lines, respectively). The gas properties of each halo are averaged within the total gas cells for each halo. The top-left panels are total galaxy group mass - gas metallicity histograms at
    $z=6$ and $15$, while the top-right panels show the same data for the subset of star-forming subhalos (those with SFR $>0$).
    In the same order, the bottom panels show the distributions of gas metallicity versus gas mass (rather than total halo mass); again, the bottom-left panels show all halos, while the bottom-right panels show only star-forming halos.  
    In all cases, the most lenient metallicity criterion ($\zgas/$\zsun\ $=0.1$) encompasses nearly $100\%$ of halos in each snapshot, while the strictest metallicity cut ($\zgas/$\zsun\ $=0.01$) includes only $94\%$ of halos and $62\%$ of star-forming halos by $z=6$. 
    \label{fig:massmet}}
\end{figure*}

Mass-metallicity histograms from Figure~\ref{fig:massmet} 
give insight on reasonable choices of BH seeding constraints for our model. We define the total gas mass and metallicity of a galaxy as that within $R_{\rm max}$, the radius at which the galaxy reaches its maximum rotational velocity.
To ensure that the subhalos selected for BH seeding are reasonably well resolved and contain a large enough gas cloud with the potential to collapse, we implement cuts on the minimum total and gas mass.  
Each model variation in minimum mass and maximum metallicity yields a
large sample of galaxies with the potential to form BHs. The question becomes: what combinations of seeding criteria produce reasonable BH populations compared to empirical data and TNG? By comparing our results with the observed BH population, we can constrain the parameter space of seeding criteria and inform future studies of BH formation and evolution.
Additionally, since TNG is known to produce good agreement with well-established local BH scaling relations, it provides a useful benchmark to compare the predictions of our SAM based seed models, particularly at higher redshifts wherein the empirical constraints are more uncertain.

As the first criteria for identifying potential BH seeding sites, we require the host galaxy to have a minimum total and gas mass. We implement total mass cuts ranging from $10^{8.5}-10^{10.5}$ \msun\ and gas mass cuts ranging from $10^7 - 10^9$ \msun. 
These values are well above the baryonic mass resolution of TNG50, $m_{\rm b} = 8.5\times10^{4}$ \msun , ensuring that the selected galaxies are well-resolved. We also explore the requirement for seeded galaxies to have nonzero star-formation rates, but in practice, we find that nearly all TNG galaxies that meet the above mass criteria are also star-forming (see Figure~\ref{fig:massmet}). 

We additionally require the potential seeding sites to have low gas metallicity.  The primordial metallicity set initially for several chemical species in TNG50 is a mass fraction of $10^{-10}$, or $10^{-8.1} Z_{\odot}$. The maximum metallicity values in our BH seeding models, set to $\zgas=10^{-1}$, $10^{-1.5}$, or $10^{-2}$\zsun, are consistent with the findings of no fragmentation occurring for gas cloud metallicities up to $Z\sim0.1$ \zsun\ for number densities as high as $10^{5}\rm{cm}^{-3}$, and where metal-line cooling does not happen effectively below $10^{-3}$ \zsun\ \citep{2009ApJ...694.1161J}. By choosing maximum metallicity values no lower than $10^{-2}$ \zsun, we also ensure that our results are well converged with resolution (see Appendix \ref{sec:res_converge}). 

Additional, complex physical processes may be involved in the formation of a BH seed that are not captured by the above seeding criteria. To account for this possibility, we also consider probabilistic seeding models with a random seeding probability $f_{\rm seed}<1$, specifically down to $f_{\rm seed}=0.01$. Each galaxy (subhalo) or galaxy group (halo) that meets all other seeding criteria in a given simulation snapshot has a probability $f_{\rm seed}$ of forming a BH in that snapshot.

Because we select BH seeding sites based solely on galaxy properties as they were computed during the actual TNG50 run and do not recompute the galaxy properties for our new SAM based seed models, there is an inherent inconsistency regarding the impact of BH feedback on host galaxies. Galaxies that have BHs within the TNG simulation (many of which will also contain BHs in our models) will experience AGN feedback effects, while galaxies that form BHs in our models but not in TNG will not experience any impact from AGN feedback. However, numerous theoretical and observational studies demonstrate that AGN feedback dominates over stellar feedback primarily in massive, low-redshift galaxies \cite[e.g.,][]{2020MNRAS.497.5292T,2019MNRAS.483.4586F,2021MNRAS.507....1V}. The primary focus of this work, in contrast, is on the formation and early growth of BHs at high redshift. Even within the high-redshift regime, massive galaxies will generally have BHs in both TNG50 and in our post-processing models. Thus, we expect this limitation to have a minimal effect on our results, and we consider this a worthwhile trade-off for the flexibility and computational efficiency of exploring a wide range of seeding models based on the TNG50 galaxy populations. The high-redshift BH seeding sites in halos have not yet undergone substantial metal enrichment through star formation, so we do not impose a minimum stellar mass criterion in order to form a BH seed, except to require that the stellar mass be nonzero. 

\subsubsection{Merger-tree Modeling of BH Populations}

To estimate the cosmic evolution of BH populations for each seeding model, we follow \texttt{SUBFIND} galaxy merger trees, each with their own unique growth histories based on seeding criteria. We trace the progenitors and descendants of galaxies that satisfy the chosen seeding criteria. The \texttt{SUBFIND} merger trees are based on the evolution of subhalos, while most seeding prescriptions in cosmological simulations rely on the properties of halos. Accordingly, the criteria for our fiducial seeding models are applied to {\em halo} properties, but to trace these seeding sites through the merger trees, we identify the central subhalo (CSH) in each halo (defined to be the most massive subhalo in a given halo). Ultimately, halo identification is then performed on the unique merger trees formed by CSH proxies. Appendix \ref{appendix:csh} indicates that this choice does not have a strong influence on our results. The use of CSH proxies does limit the models from seeding BHs in satellite galaxies within halos, but in practice, the population of TNG satellites that meet the model seeding criteria and also have BHs is small (see Figure \ref{fig:host_comparison}).

Regardless, this approach does necessitate the simplifying assumption of one BH per halo, such that when two galaxies merge and each contains a BH seed, we assume the BHs also promptly merge. 
This treatment gives a lower limit on the BH number densities and the merger timescales for each seeding model. It is a rough approximation over the course of descendant evolution, because BH merger timescales can be several Gyr if binary inspiral is inefficient. (Interestingly, recent analysis of the PTA evidence for a stochastic GW background suggests that short inspiral timescales are favored by the data \citep{2023ApJ...952L..37A}. These early results are still too tentative to provide a robust justification for our simplifying assumption, however.) A detailed study (e.g., using different BH growth or dynamical friction models) aimed at examining BH merger rates or LISA event rates  would warrant a more realistic treatment of BH binary inspiral timescales. Gravitational recoil would also be important to consider BH retention within the galaxies. We plan to focus on these details in future studies. 

\subsubsection{Modeling BH Growth}
\label{sec:SRMs}

As noted in \S\ ~\ref{sec:seedingsites}, our post-processing scheme for seeding BHs and tracing them through galaxy merger trees provides BH number densities and occupation fractions, but it does not allow for BH masses and accretion rates to be obtained directly from the simulation. In order to compare our seeding model results with empirical measurements of BH mass functions and mass density evolution with redshift, we employ a simple prescription to assign masses to the BHs when they form and as they evolve through time.

Specifically, the mass of each BH is assumed to be a constant fraction of the galaxy's total stellar mass at each point in time. We choose a mass fraction of $10^{-3}$, motivated by empirical constraints on the BH mass-stellar bulge mass relation, fitted from early and late-type galaxies \cite[e.g.,][]{2013ApJ...764..184M}. Since the BHs are assumed to merge when the galaxies merge within the merger trees, this means that the assumed BH mass depends on the combined total stellar mass of the merged galaxy. Owing to the poor constraints on the BH-bulge relation at high redshift and the approximate nature of our approach, we rely on the total stellar mass rather than performing a kinematic decomposition of each galaxy's stellar bulge and disk components, and we do not explore the impact of scatter in the BH-bulge relations. 
This enables us to make quick, rough BH mass estimates and determine which seeding models produce BH populations in reasonable agreement with empirically derived BHMFs and mass densities. We note also that different empirical measurements of these quantities vary significantly, especially at high-redshift \citep{2008MNRAS.388.1011M,2009ApJ...690...20S,2010ApJ...725..388C,2020MNRAS.495.3252S}. In future work, we plan to undertake a more detailed exploration of BH mass growth prescriptions for our SAMs.

\subsubsection{Parameter space of BH seeding SAMs}
\label{sec:Parameter_space}

We consider a wide selection of seed models that can be divided into two categories based on whether we are systematically varying the minimum threshold for total halo mass ($\mtot$) or for halo gas mass ($\mgas$). Each model also includes a maximum threshold for the average gas metallicity of halos~($\zgas$), and a probability of seeding $f_{\rm seed}$. For each ($\mtot$, $\mgas$) pair, we consider six different models with $\zgas/$\zsun$~ = 10^{-1}$, $10^{-1.5}$, or $10^{-2}$ and $f_{\rm seed} = 0.01$ or $1$.

\begin{table*}
\begin{center}
\begin{tabular}{cccccc}
\hline

Model type & Model names & $\mtot$ & $\mgas$ & $\zgas$ & $f_{\rm seed}$\\
  &  & [log$_{10}$ \msun] & [log$_{10}$ \msun] & [log$_{10}$ \zsun] & - \\
\hline
Varying $\mgas$ & \texttt{mgas[7,8,9]\_Z[0.01,0.03,0.1]} & 8.0 & (7.0, 8.0, 9.0) & (-2.0, -1.5, -1.0) & (0.01, 1.0) \\
Varying $\mtot$ & \texttt{mtot[8.5,9.5,10.5]\_Z[0.01,0.03,0.1]} & (8.5, 9.5, 10.5) & 7.0 & (-2.0, -1.5, -1.0) & (0.01, 1.0) \\
\hline

\end{tabular}
\end{center}

\caption{Summary of semi-analytic BH seeding models used in this work. For each model type (``varying $\mgas$" or ``varying $\mtot$"), we consider three values of the relevant mass threshold (while keeping the other mass threshold fixed), as well as three values of $\zgas$ and two values of $f_{\rm seed}$. Our SAM suite therefore includes 36 distinct BH seeding models. Model names specify the variable mass threshold and the metallicity threshold: \texttt{mgas*\_Z*} or \texttt{mtot*\_Z*}. $f_{\rm seed}$ is not included in the nomenclature, as all results are presented as a range of values when $f_{\rm seed}$ is varied from 0.01 to 1.}
\label{list_of_seed_models}
\end{table*}

We label these types of seed models as \texttt{mgas*\_Z*} and \texttt{mtot*\_Z*} respectively, where the asterisks denote the appropriate values for each parameter. For example, \texttt{mgas7\_Z0.1} refers to a seed model with $\mgas=10^7$~\msun\ and $\zgas=10^{-1}$~\zsun. Similarly, \texttt{mtot8.5\_Z0.03} refers to a model with $\mtot=10^{8.5}$~\msun\ and $\zgas=10^{-1.5}$~\zsun. In all of our figures, results are presented for each ($\mgas$, $\mtot$, $\zgas$) combination as a range of values spanning $f_{\rm seed} = 0.01$ - $1$. Thus, the value of $f_{\rm seed}$ is not included in the model nomenclature. These models and their nomenclature are summarized in Table~\ref{list_of_seed_models}.

\section{Results}
\label{sec:results}
\subsection{Mass-Metallicity Relations of High Redshift Halos}
\label{sec:massmet_relations}
 
 \begin{figure*}
    \centering
    \includegraphics[width=\textwidth]{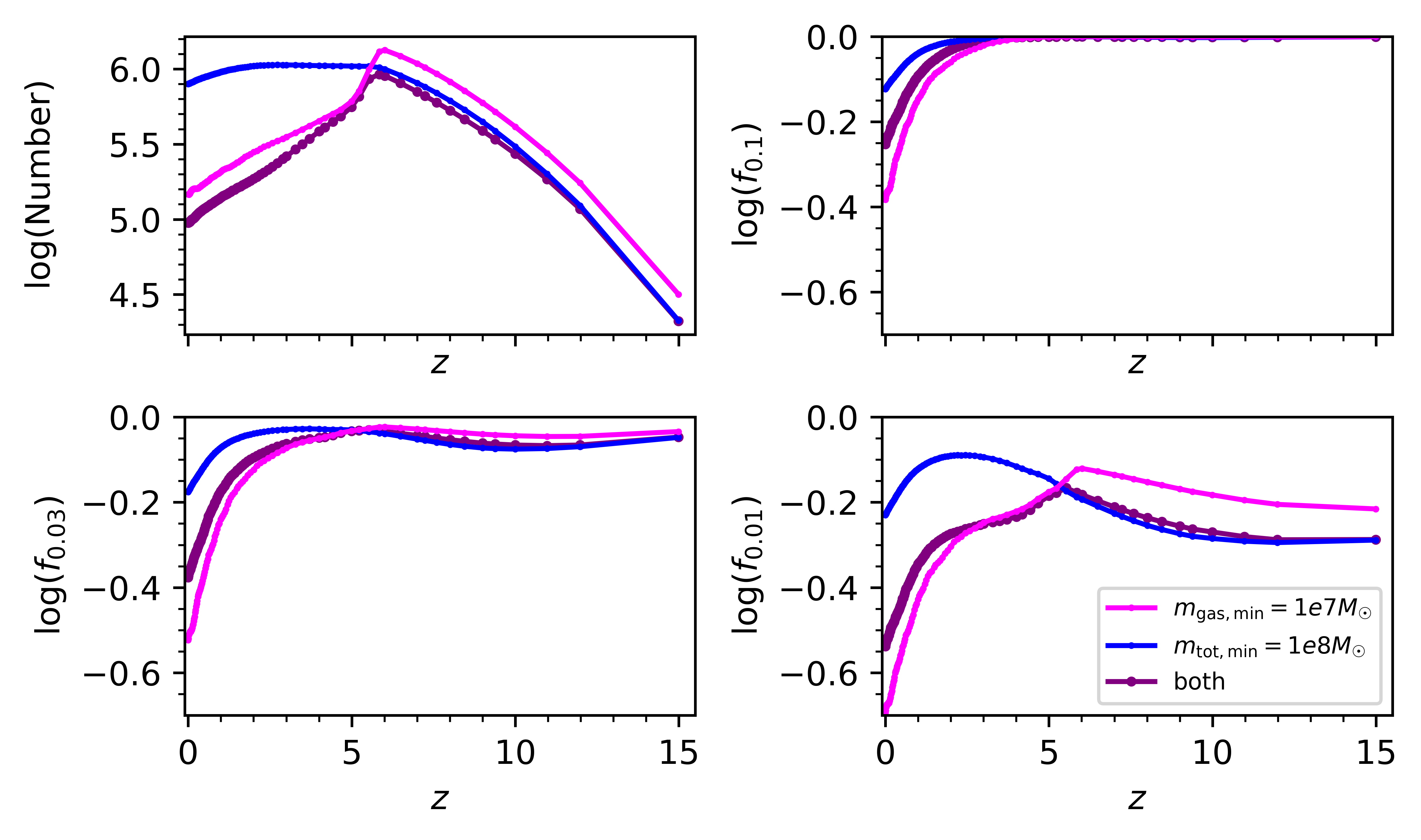}
\caption{The evolution of metal-poor halo sub-populations is shown. In the top left panel, we show the total number of halos above a minimum mass of $m_{\rm{tot,min}}=10^{8}$ \msun\ (in blue), $m_{\rm{gas,min}}=10^{7}$ \msun\ (in magenta), or both (in purple). All other panels show the fraction of these halos that are metal-poor as defined via their average gas-phase metallicity, using the same color scheme. The top right panel shows the fraction of halos that satisfy $\zgas/$\zsun~$=0.1$, which we denote as $ f_{0.1}$, and the bottom left and bottom right panels show the fractions with $\zgas/$\zsun~$= 10^{-1.5}$ or $10^{-2}$, denoted as $f_{0.03}$ and $f_{0.01}$, respectively. We see that $f_{0.1}$ remains very high until $z\sim4$ and then declines as halo enrichment proceeds towards cosmic noon. Stricter metallicity cuts show similar trends but are also modulated by the increase in total number of halos up to $z\sim6$.
\label{fig:H_frac_Z}}
\end{figure*}

In Figure~\ref{fig:massmet}, we examine the distributions of key galaxy properties at high redshift, focusing on gas metallicity versus gas mass and halo mass in the $z=15$ and $z=6$ snapshots. We study these galaxy populations to inform the different mass cuts and metallicity cuts that we plan to apply in our SAM based seed models, as summarized in Section \ref{sec:Parameter_space} and Table \ref{list_of_seed_models}.  Low-metallicity galaxies with $\zgas/$\zsun $=10^{-2}, 10^{-1.5},\rm{or}\ 0.1$
make up the majority of hosts at both redshifts. Nearly $100\%$ of halos meet the most lenient metallicity cuts $\zgas/$\zsun~$=0.1$ at $z\sim15$ and $z\sim6$. 
Considering the strictest metallicity cuts $\zgas/$\zsun~$ = 10^{-2}$, and discounting halos with no gas at all, the proportions of halos that satisfy this criterion decrease from $97\%$ to $94\%$ from $z=15$ to $6$. The same fraction decreases from $92\%$ to $62\%$ for the star-forming population.

In Figure~\ref{fig:H_frac_Z}, we investigate the fraction of low-metallicity halos that satisfy the most lenient minimum mass criterion in our seed models: $m_{\rm{gas}} > 10^{7}$ \msun\ and/or $m_{\rm{tot}} > 10^{8}$ \msun. Since nearly all of these galaxies exhibit active star formation, we have excluded an additional star-formation criterion from Figure~\ref{fig:H_frac_Z}. In the absence of any metallicity criteria (top left panel of Figure~\ref{fig:H_frac_Z}), the total number of halos meeting these minimum mass criteria grows from a few $\times 10^4$ at $z=15$ to $\gtrsim 10^6$ by $z=6$.  Star formation, feedback processes, and mergers subsequently reduce the number of halos meeting the gas mass criterion after $z\sim6$. By $z=0$, there are roughly $8\times10^{5}$ halos that satisfy $m_{\rm tot}>10^{8}$ \msun, $1.5\times10^{5}$ halos that satisfy $m_{\rm gas}>10^{7}$ \msun, and $9.5\times10^{4}$ halos that satisfy both criteria. 

The remaining three panels in Figure~\ref{fig:H_frac_Z} show the fraction of these halos that meet not only the 
specified mass cuts but also satisfy the maximum metallicity cuts $\zgas/$\zsun~$=10^{-2},10^{-1.5},\rm{or}\ 0.1$, 
denoted as $f_{0.01},f_{0.03},\rm{and}~f_{0.1}$, respectively. The top right panel shows that nearly all of these halos have $\zgas/$\zsun~$=0.1$, through the epoch of reionization. It is only at redshifts below $z\sim4$ that the metal-poor fraction noticeably declines, as the Universe approaches the peak of star-forming activity at ``cosmic noon."  
For the population that satisfies both mass cuts, $f_{0.1}$ goes from nearly $100$\%\ at $z\sim15$ to $56\%$ at $z\sim0$.

With a stricter metallicity cut of $\zgas/$\zsun~$=10^{-1.5}$, we see broadly similar behavior with some minor differences. Roughly $90$\%\ of these halos are below 
this enrichment level at $z\sim15$. We also see a slight temporary dip in the fraction of metal-poor halos between $z=15$ and $z\sim6$, 
owing to the interplay between halo enrichment via star formation and the steady increase in the total number of metal-poor halos meeting the mass cuts. Below $z\sim6$, the number of halos levels out and eventually declines due to mergers, while the number of halos above $m_{\rm gas,min}$ sharply declines owing to a burst of star formation and feedback. After this point, continued metal enrichment steadily decreases the fraction of metal-poor halos. 
These trends are starker for the lowest metallicity threshold $\zgas/$\zsun~$=10^{-2}$. 
Only about $52$\%\ of halos exceeding both mass thresholds lie below $\zgas/$\zsun~$=10^{-2}$ 
at $z=15$, and by $z=0$ this metal-poor fraction is $29$\%.
 
\begin{figure*}
\centering
\includegraphics[width=0.6\textwidth]{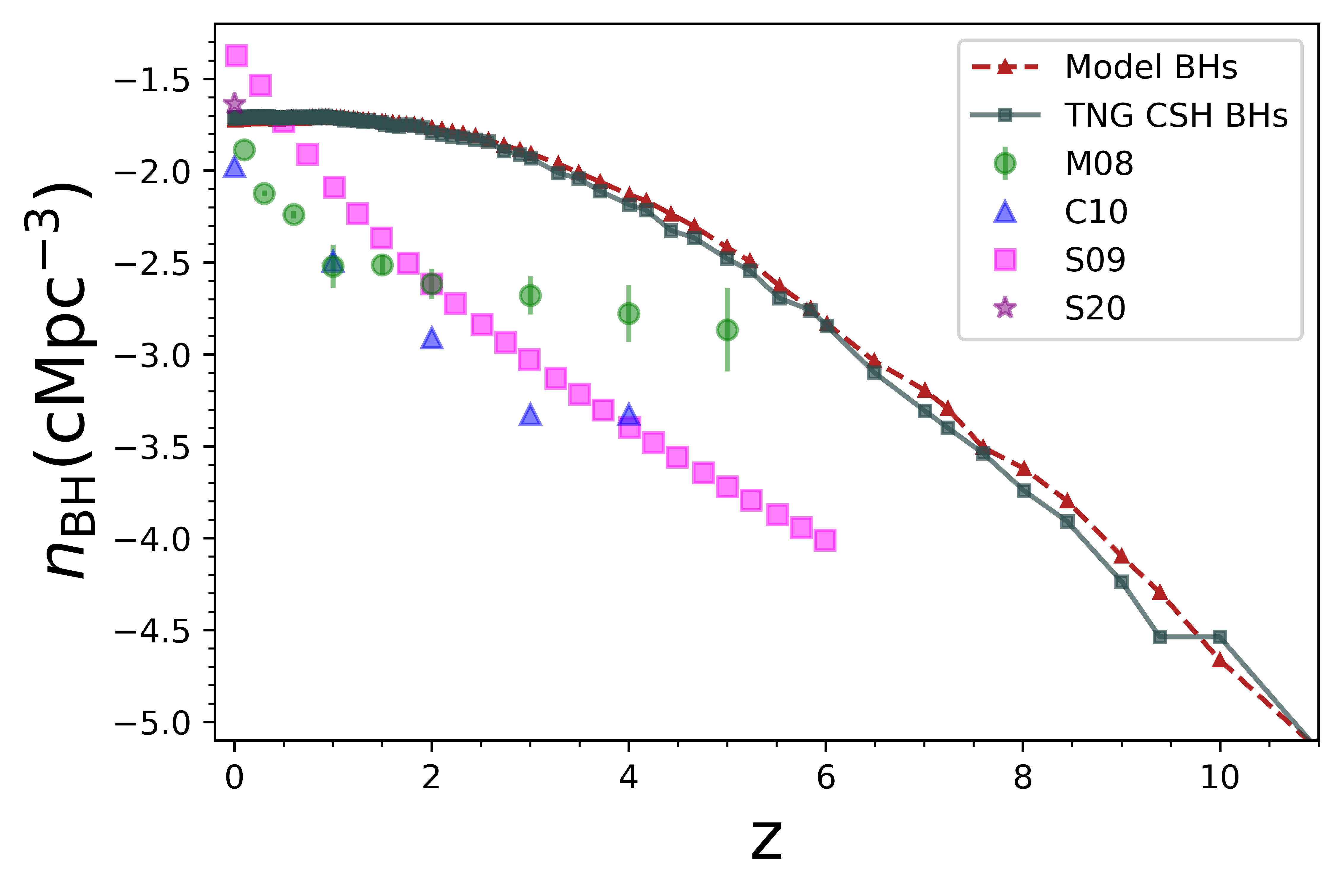}
\caption{We apply the TNG halo criterion $5\times10^{10}$ \msun \hinv and choose a CSH proxy for these eligible halos for our halo models, assuming one BH per halo. We compare the halo model results to TNG CSH BHs due to our model choice of CSH proxies. The number density ($n_{\rm{BH}}$) of BHs (in units of 
comoving Mpc$^{-3}$) in our TNG model (in the dashed red line) comes close to that of TNG CSHs (in the solid dark gray line) at all redshifts and to within $4\%$ at $z=0$. $n_{\rm{BH}}$ from the empirical studies M08 (green circles), C10 (blue triangles), S09 (magenta squares), and S20 (purple star) are shown (see full references in \S~\ref{tng_vs_model}). At $z\gtrsim0.5$, TNG predicts higher number densities than observations, but it lies squarely in the middle of the empirical data at $z=0$. 
\label{fig:tng_BHs_z}}
\end{figure*}

\begin{figure}
\centering
\includegraphics[width=\columnwidth]{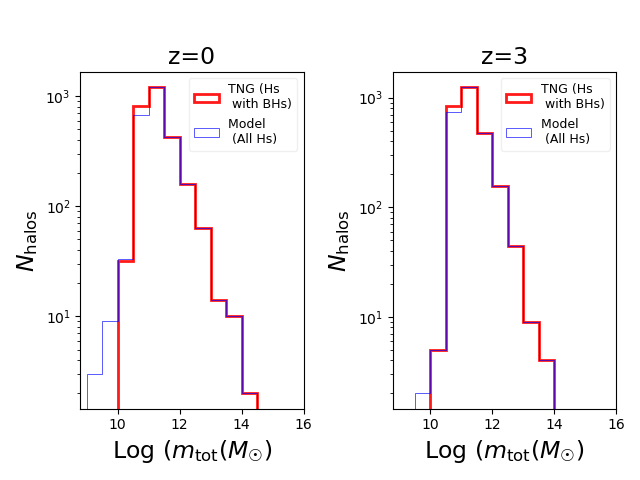}
\caption{The total mass distributions at $z=0$ and $z=3$ are shown for halos that have BHs in TNG (red) versus the halos hosting BHs in our TNG-analogue model (blue). Both histograms show close agreement between host masses from the model and TNG halos.
\label{fig:tng_model_histos}}
\end{figure}

\subsection{SAM verification: Reproducing the TNG BH population}
\label{tng_vs_model}

Before we attempt to explore the different physically motivated, SAM-based seed models summarized in Section \ref{sec:Parameter_space}, we first verify that our approach can successfully reproduce the actual TNG results when the TNG seeding criterion is applied. We 
impose a minimum halo mass of $5\times10^{10}$ \msun \hinv, consistent with the TNG BH seeding criterion. Figure~\ref{fig:tng_BHs_z} compares our TNG-analogue semi-analytic seeding prescription to the true number density of BHs within the CSHs in the TNG simulation. Because of the CSH proxies used in the models, number density evolution is compared with that from TNG CSH BHs, but there is little difference between the halo and CSH model results; this means that the population of satellite galaxies in TNG that meet the model seeding criteria and host BHs is small (see Appendix~\ref{fig:host_comparison}). 
The model agrees well with the actual number density of BHs in TNG at all redshifts; at $z=0$, the model agrees with TNG to within $4$\%.

In Figure~\ref{fig:tng_BHs_z}, model results are also compared with empirical BH number densities \citep[][hereafter referred to as M08, S09, C10, and S20, respectively]{2008MNRAS.388.1011M,2009ApJ...690...20S,2010ApJ...725..388C,2020MNRAS.495.3252S}.
M08 and C10 both use the BHMF continuity equation but make different assumptions about the growth of the BHs. M08 empirically determine the Eddington ratio distribution by coupling the empirical BH mass function and X-ray luminosity function with fundamental relations between three different accretion mode observables, while C10 assumes a power-law Eddington ratio distribution.
S09 models AGN and SMBH populations under the assumption that the BHMF grows at the rate implied by the observed luminosity function. S20 give updated constraints on the bolometric quasar luminosity function from observations from the past decade with an updated quasar SED model and bolometric and extinction corrections. At $z\sim0$, these studies predict BH number densities ranging from $1.3-4.2 \times 10^{-2}$ cMpc$^{-3}$; 
the TNG $z=0$ number density of $n_{\rm BH} = 1.96 \times 10^{-2}$ lies in the middle of these values. 

Notably, there are substantial discrepancies between the different empirical constraints on the BH number density, which increase at higher redshift. There are also significant discrepancies between the TNG and the empirically estimated BH number densities, especially at high redshift. Previous studies have similarly found that although the low-redshift TNG QLFs 
and the $z=0$ BHMFs agree reasonably well with observations \citep{2015MNRAS.452..575S,2018MNRAS.479.4056W}, TNG overpredicts the high-redshift QLF \citep{2018MNRAS.479.4056W}. Other simulations using similar physical models have also been found to overpredict the bright end of the AGN luminosity function at high redshift \citep{2021MNRAS.507.2012B}. However, high-redshift quasar statistics remain incomplete and poorly constrained, particularly at the faint end of the luminosity function. This creates large uncertainties in the BHMF at early times, especially at the low mass end. 

JWST has already uncovered substantial new populations of AGN at high redshifts \citep[e.g.,][]{2023ApJ...942L..17O,2023arXiv230308918L,2023arXiv230512492M,2023arXiv230200012K} and will transform our understanding of the high-redshift AGN luminosity function in the coming years. Advances in theoretical models of high-redshift BH populations will be crucial for interpreting this new wealth of data from JWST, and in preparation for LISA observations of the high-redshift GW Universe. The large BH seed masses used in many simulations ($\sim10^6 M_{\odot}$ in TNG) likely contribute to overestimation of the low mass end of the BHMF at high redshift, but at the same time, observational constraints on low-mass, high-redshift BHs are highly incomplete. 
This is precisely one of the issues that our present work addresses by modeling BH populations with lower seed masses and a much wider range of seeding criteria.

The host mass histograms in the left panels of Figure~\ref{fig:tng_model_histos} at $z=0$ and $z=3$ show that not only does the total number of BHs agree well between our semi-analytic model and TNG, but also the distribution of host halo masses. Note that in both cases, a tail of BH host masses extends below the minimum required halo mass for BH seeding in TNG ($5\times 10^{10}$ \msun\hinv), especially at $z=0$. These are galaxies that have lost mass over time via tidal stripping.

\begin{figure*}
\centering

\includegraphics[width=0.32\textwidth]{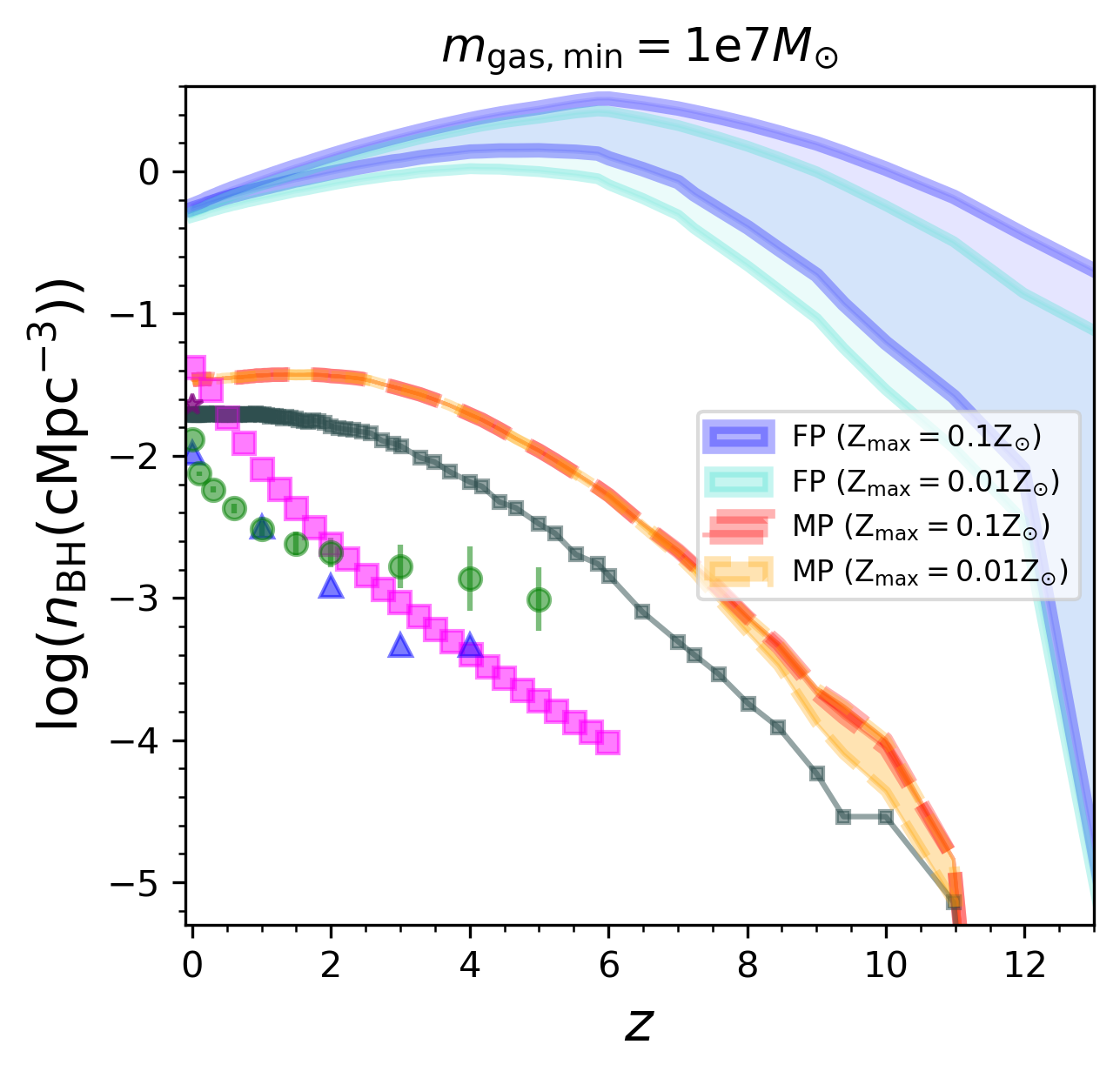}
\includegraphics[width=0.32\textwidth]{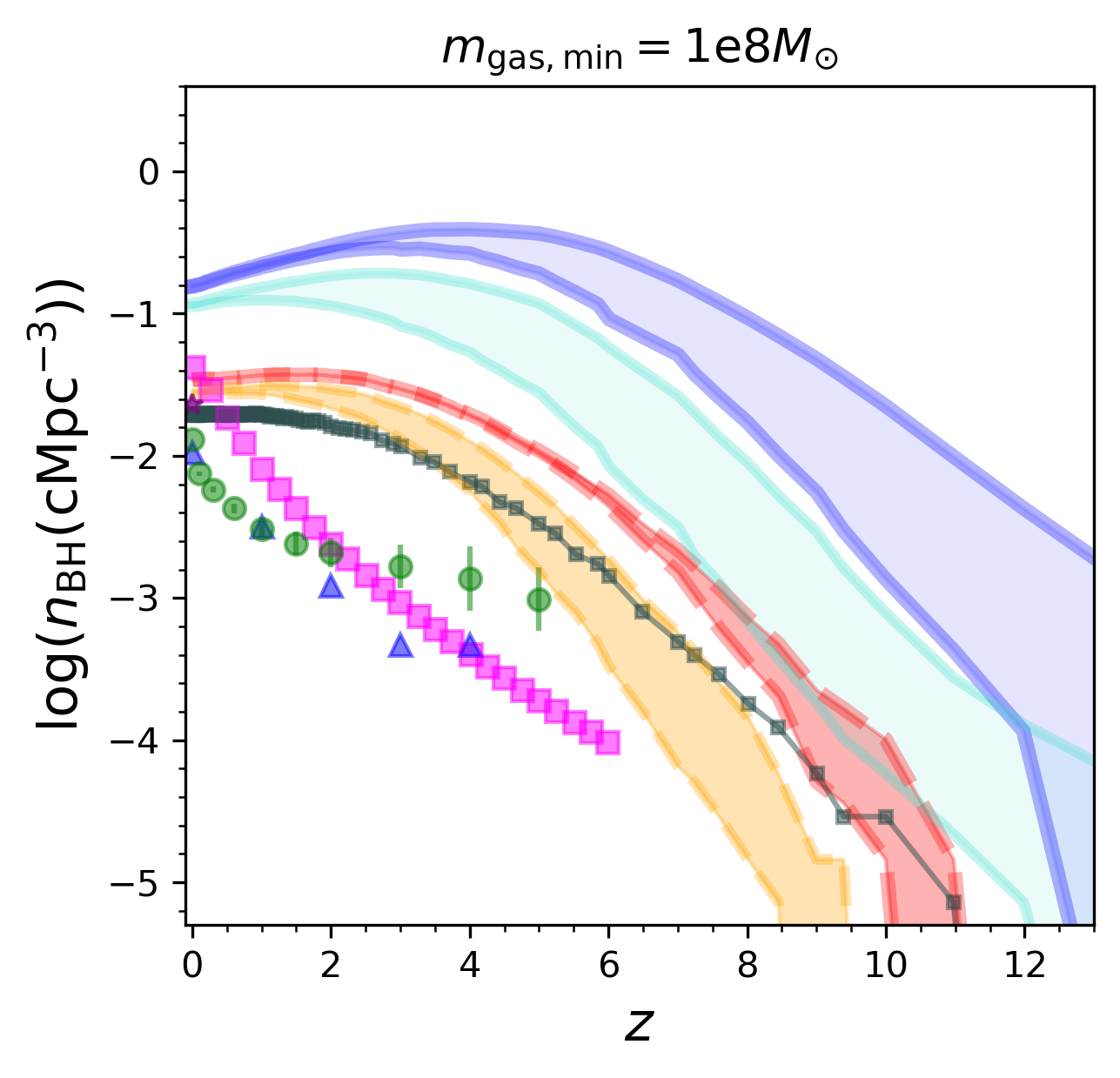}
\includegraphics[width=0.32\textwidth]{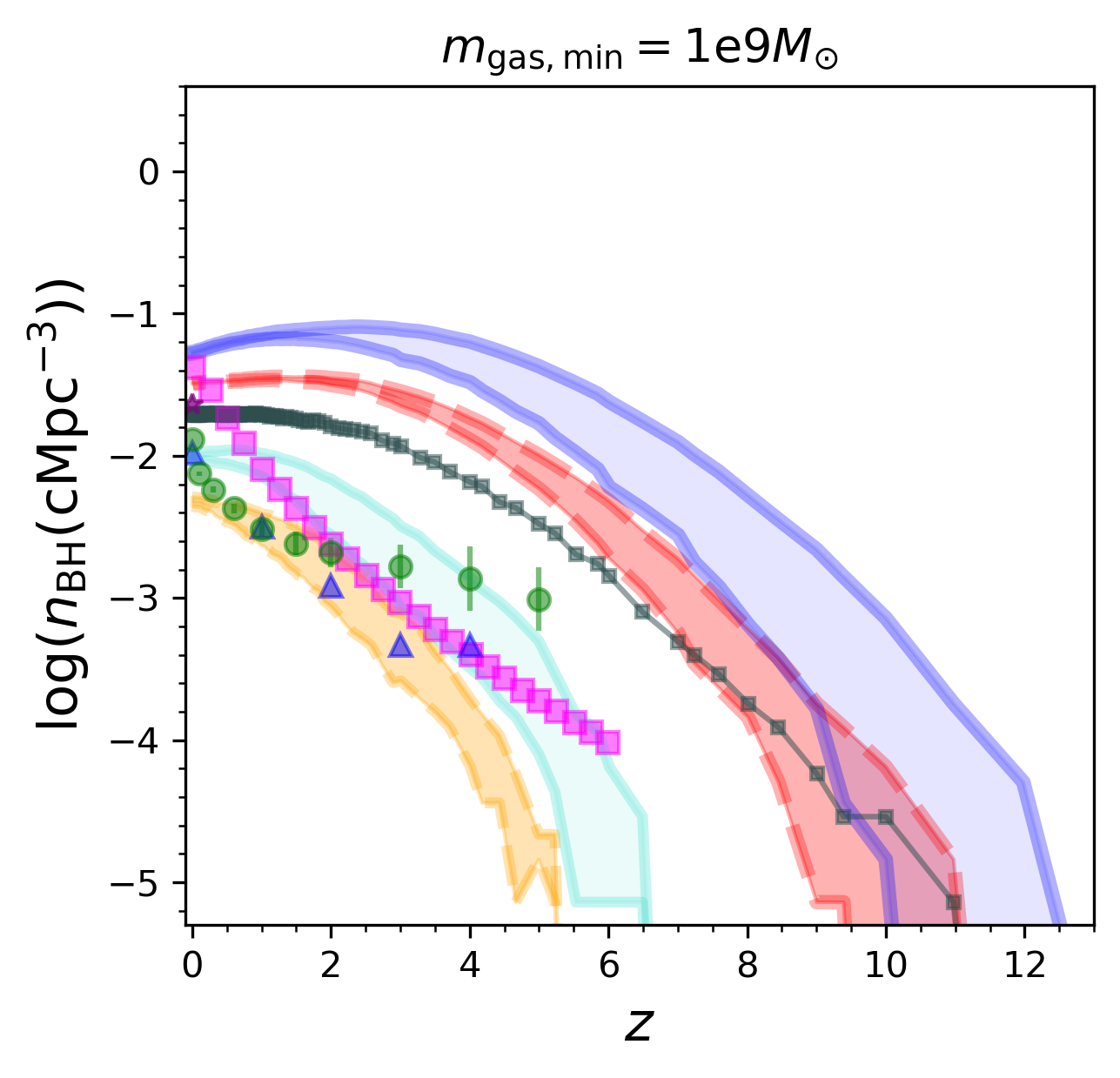}

\includegraphics[width=0.32\textwidth]{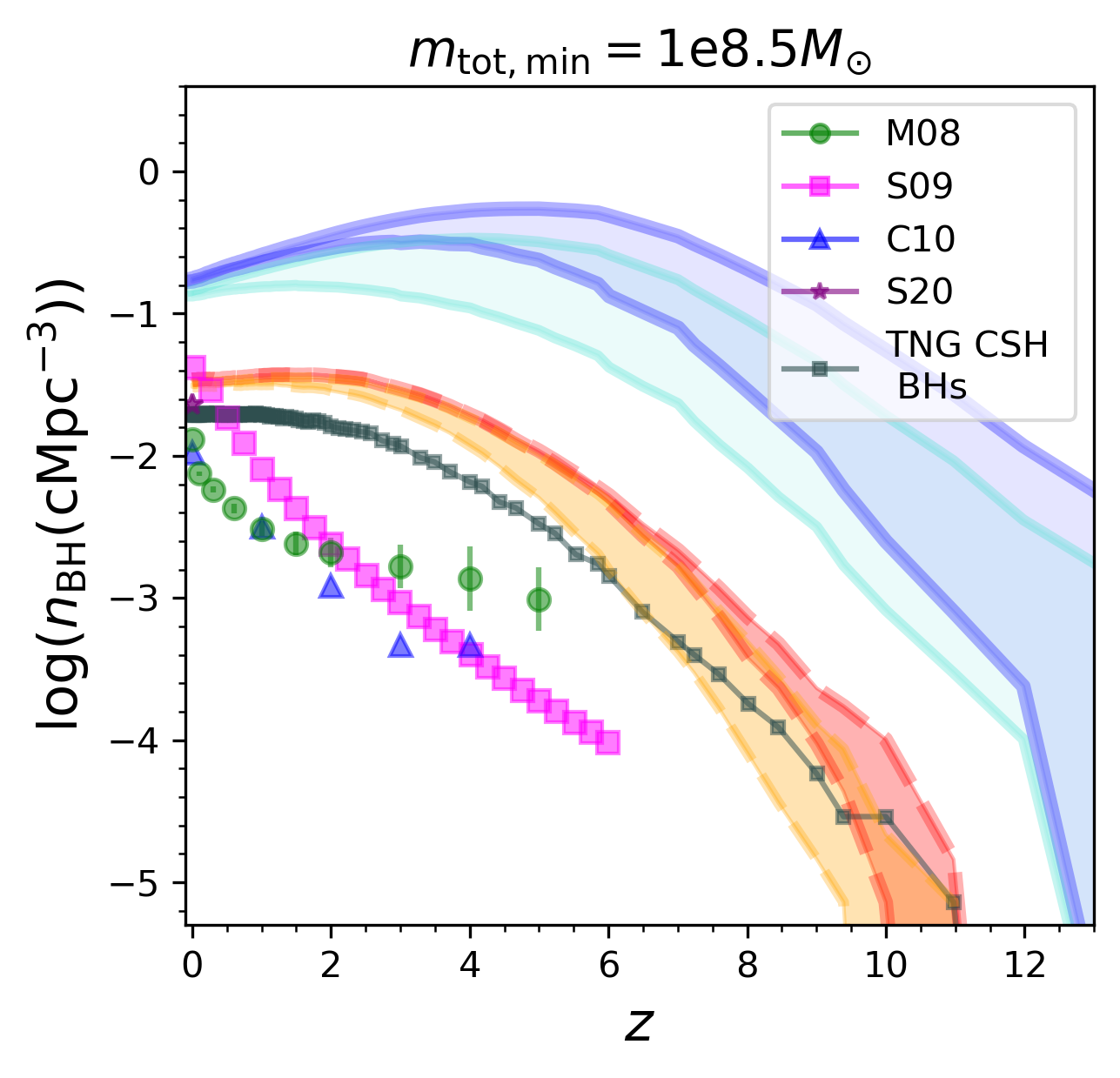}
\includegraphics[width=0.32\textwidth]{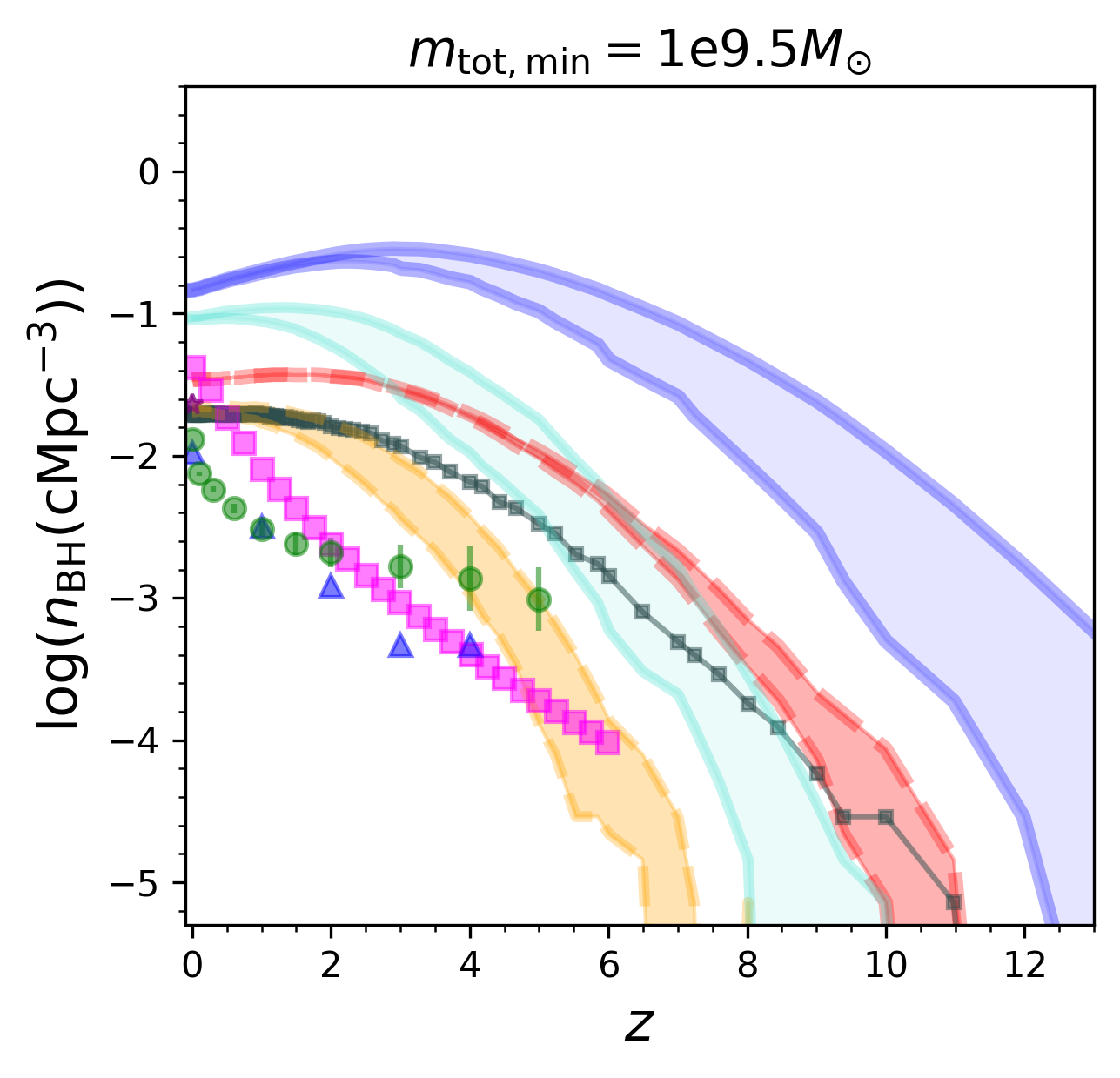}
\includegraphics[width=0.32\textwidth]{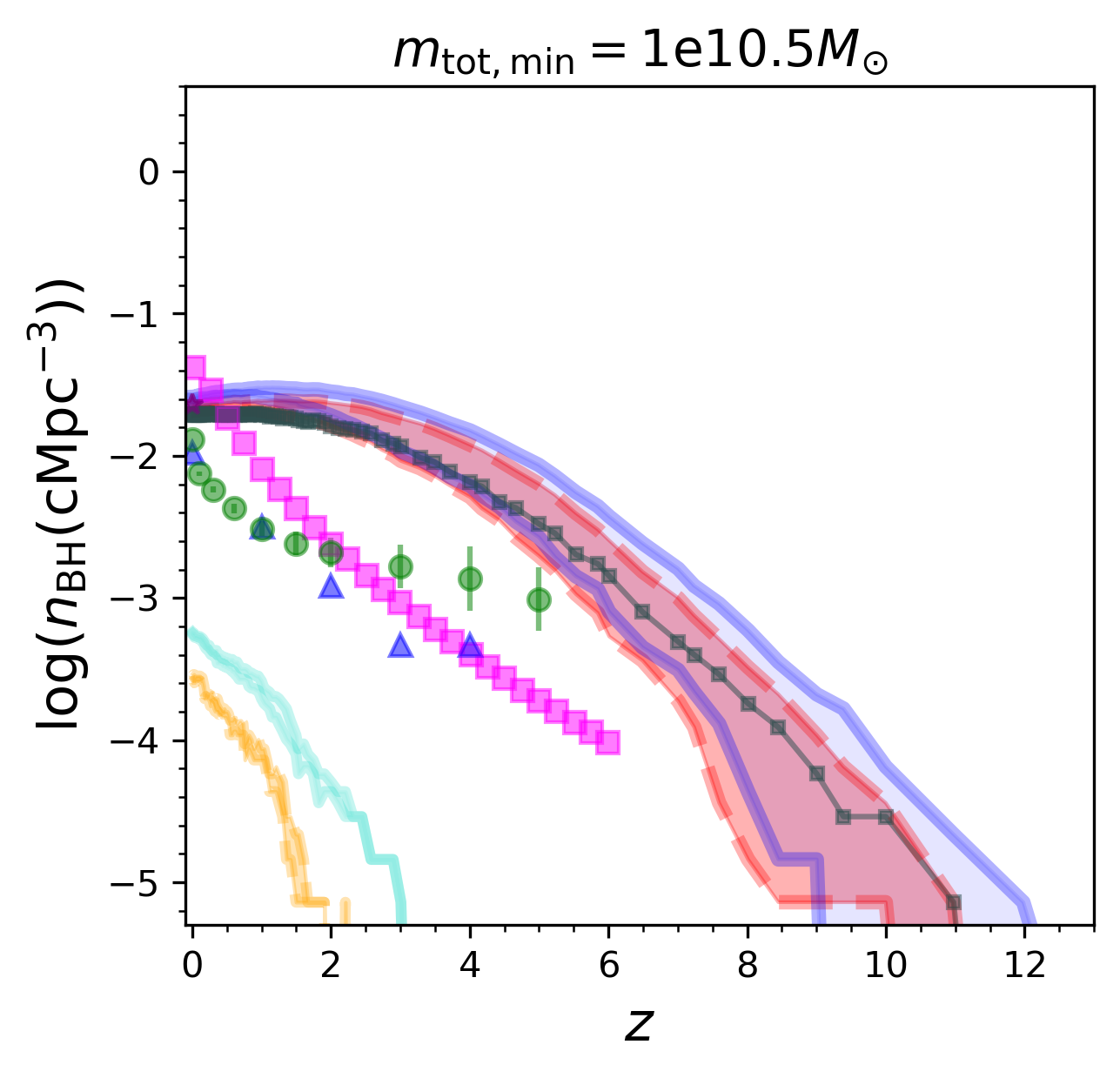}

\caption{Comoving BH number densities, $n_{\rm{BH}}$, are shown versus redshift for fiducial halo models. FP and MP results are shown in cool and warm-colored transparent shaded regions, respectively. The lower and upper limits of the shaded regions correspond to probabilistic seeding fractions between $0.01$ and $1$. The models differ in gas mass, host mass, and metallicity criteria ($\zgas = 0.1$\zsun ~ \rm{and}~$10^{-2}$\zsun) (in red and gold, respectively, for the MP, and in blue and turquoise, respectively, for the FP). The top panels correspond to varying-$\mgas$ models and the bottom panels correspond to varying $\mtot$. The parameters were chosen systematically and not with the intent of producing the closest fit. Several different SAMs span number densities that agree with results from TNG and AGN observations (M08, C10, S09, and S20). 
At high redshifts, BHs $<10^{5}$ \msun\ are the most significant contributors to the number densities by factors of $\sim10-100$. This underscores the importance of LISA's capabilities to detect these systems at high redshift.
\label{fig:n_BH_intro.png}}
\end{figure*}

\begin{figure*}
\centering

\includegraphics[width=0.32\textwidth]{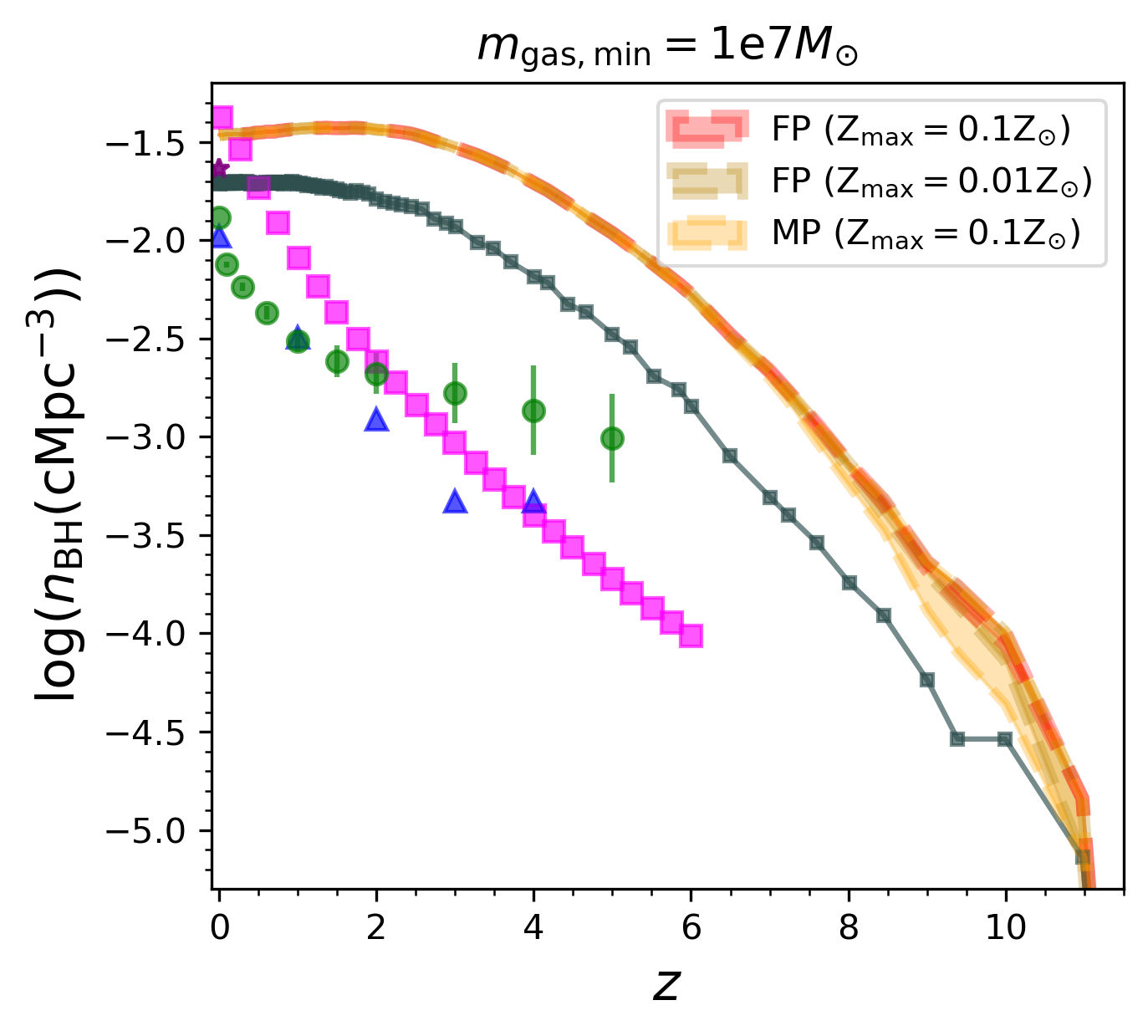}
\includegraphics[width=0.32\textwidth]{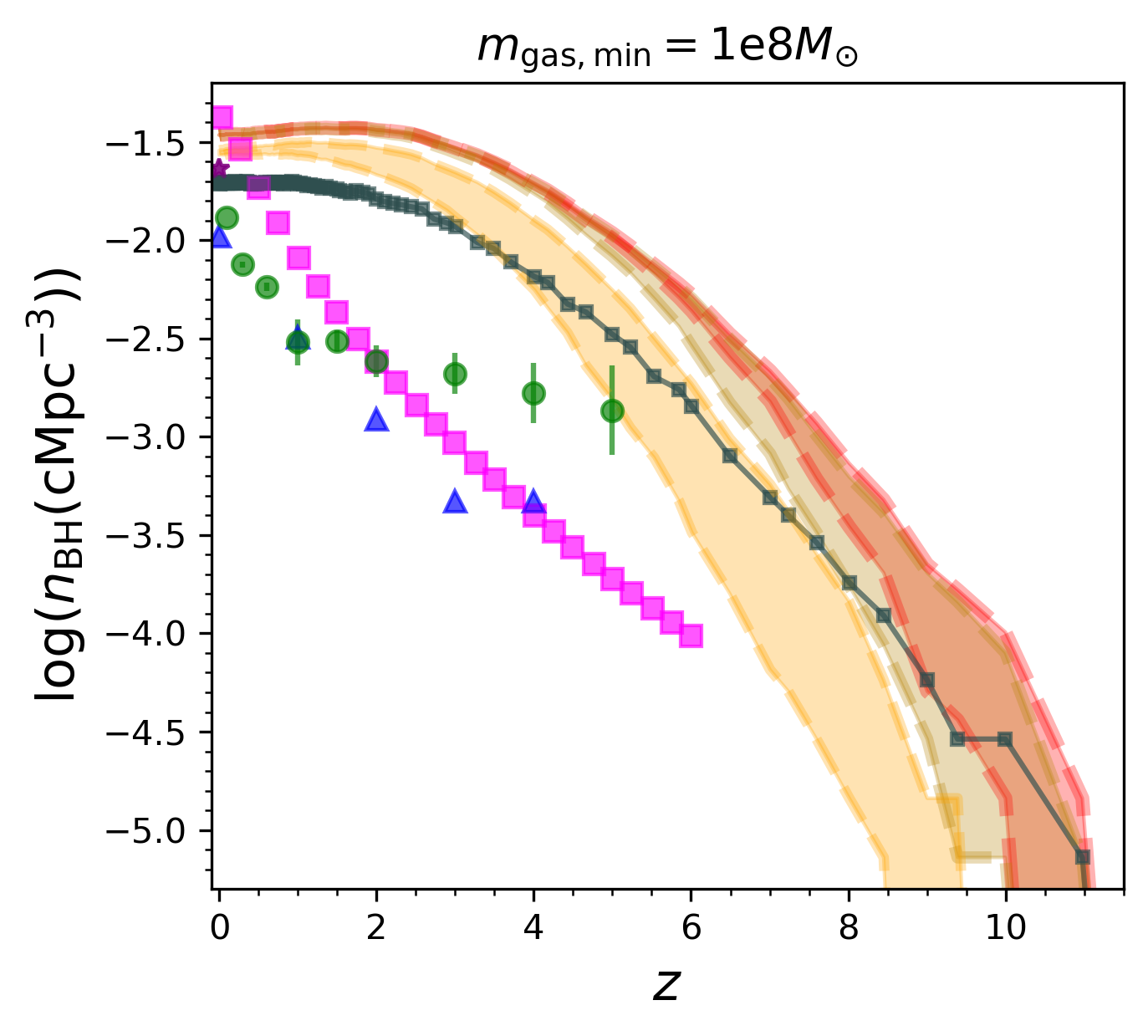}
\includegraphics[width=0.32\textwidth]{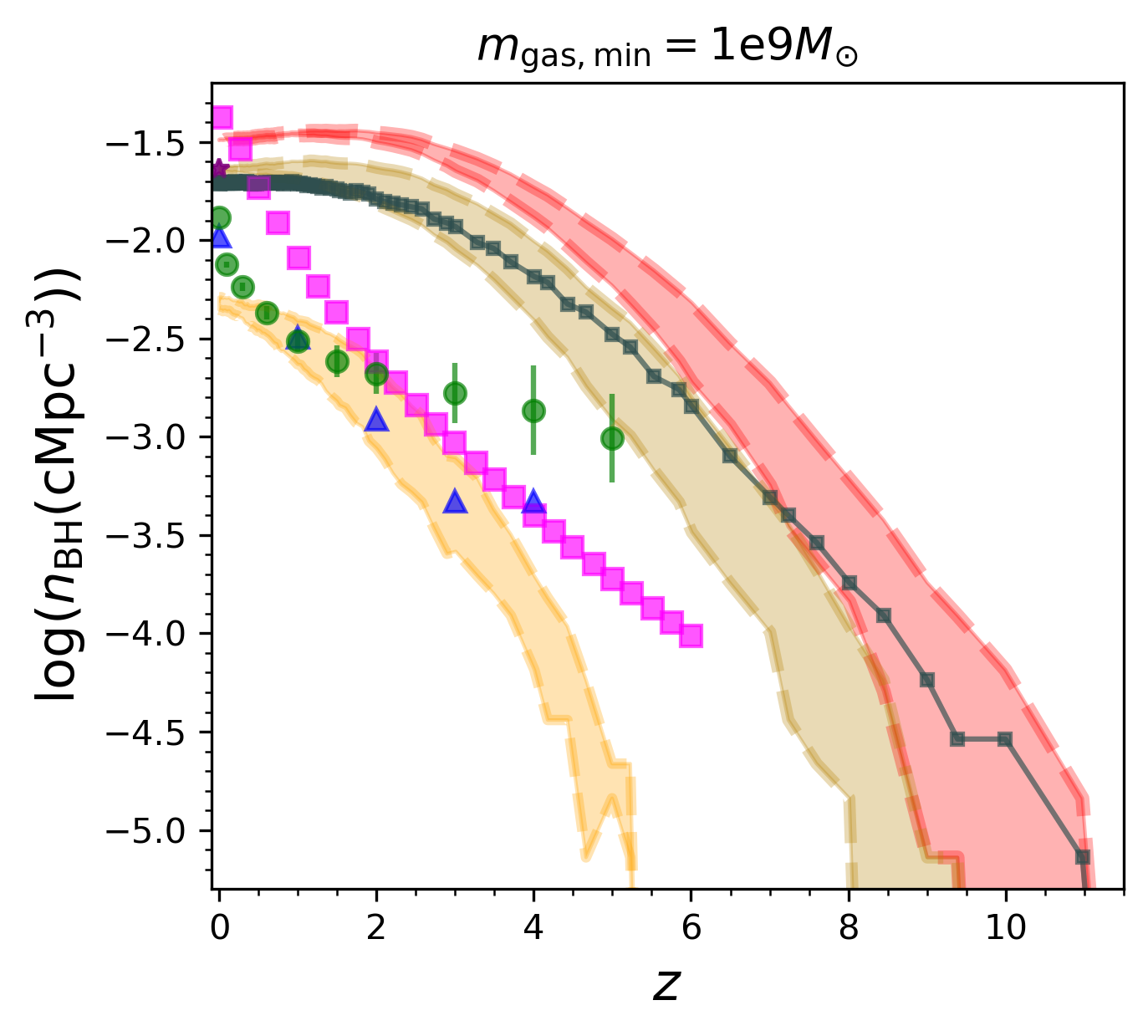}

\includegraphics[width=0.32\textwidth]{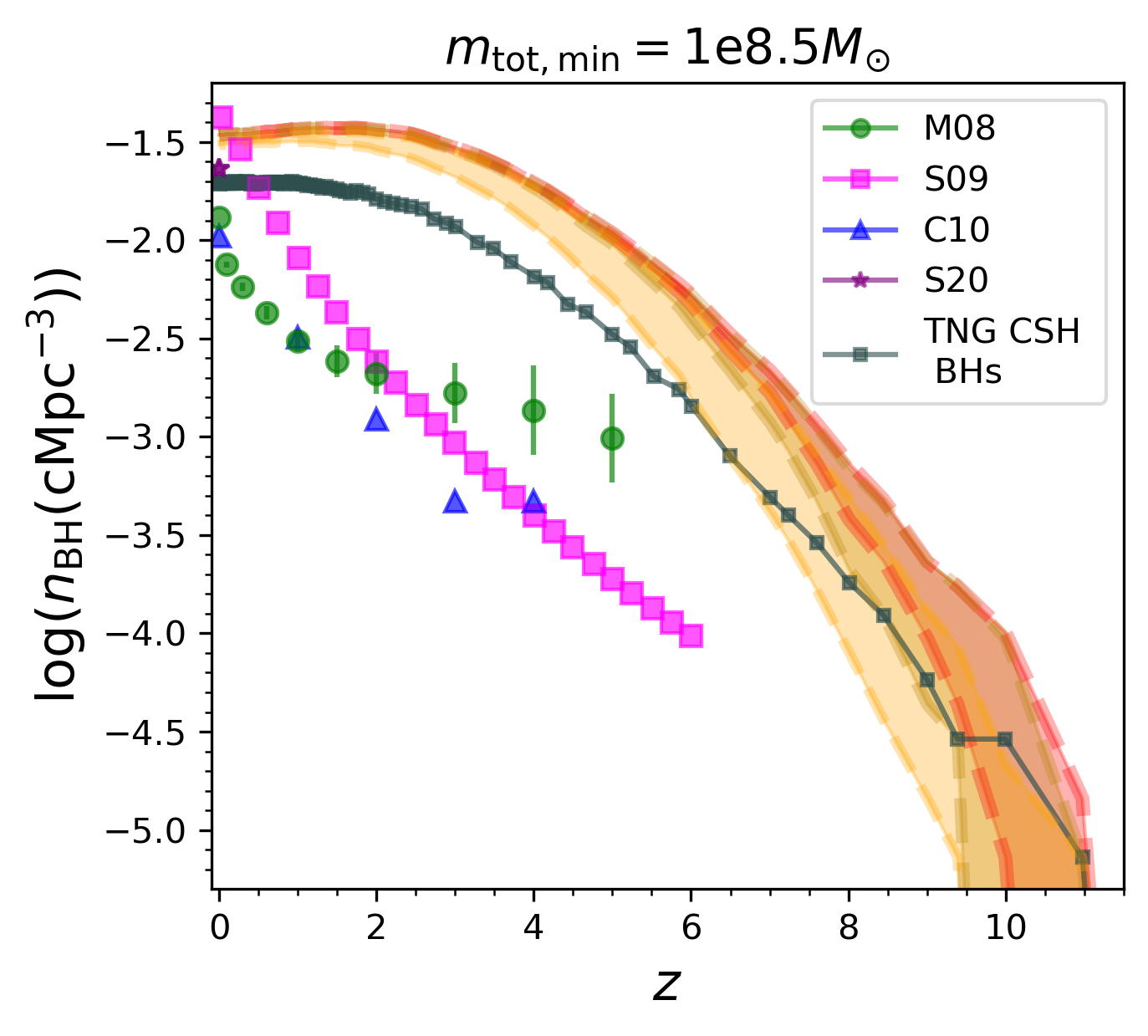}
\includegraphics[width=0.32\textwidth]{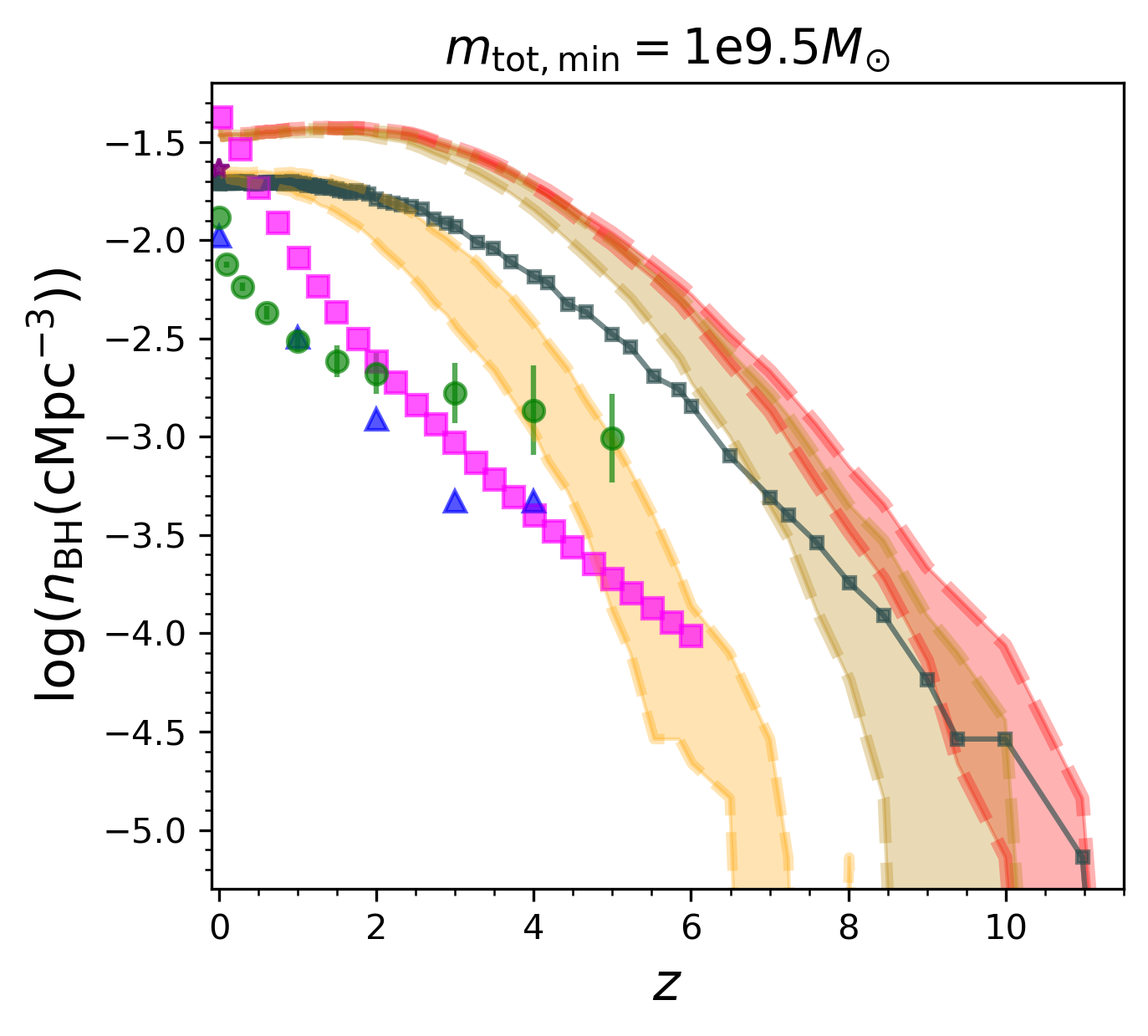}
\includegraphics[width=0.32\textwidth]{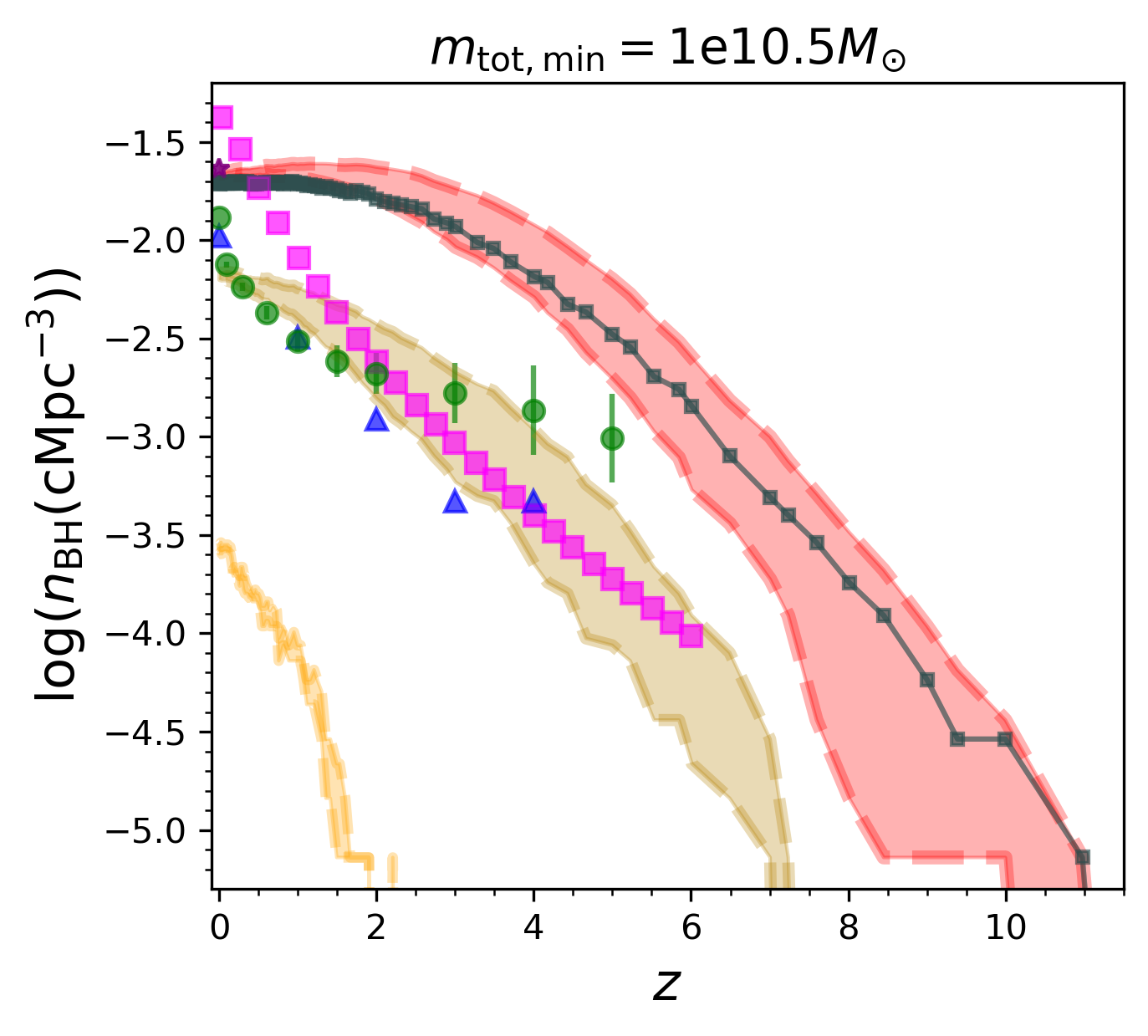}

\caption{Comoving BH number density evolution is shown in the same format as Figure~\ref{fig:n_BH_intro.png}, except only the MP model results are shown (i.e., all results include only BHs $>10^{5}$ \msun). In addition, SAMs with intermediate host metallicity thresholds of $\zgas/$\zsun~$= 10^{-1.5}$ are shown in the tan shaded region. Numerous models show reasonable agreement with empirical constraints, within the considerable observational uncertainties. The models spanning the empirical space illustrate their capabilities to explore more realistic seed mass variations by seeding in lower-mass hosts than the TNG halo seed mass threshold. 
\label{fig:number_densities.png}}
\end{figure*}

\begin{figure*}
\centering

\includegraphics[width=0.32\textwidth]{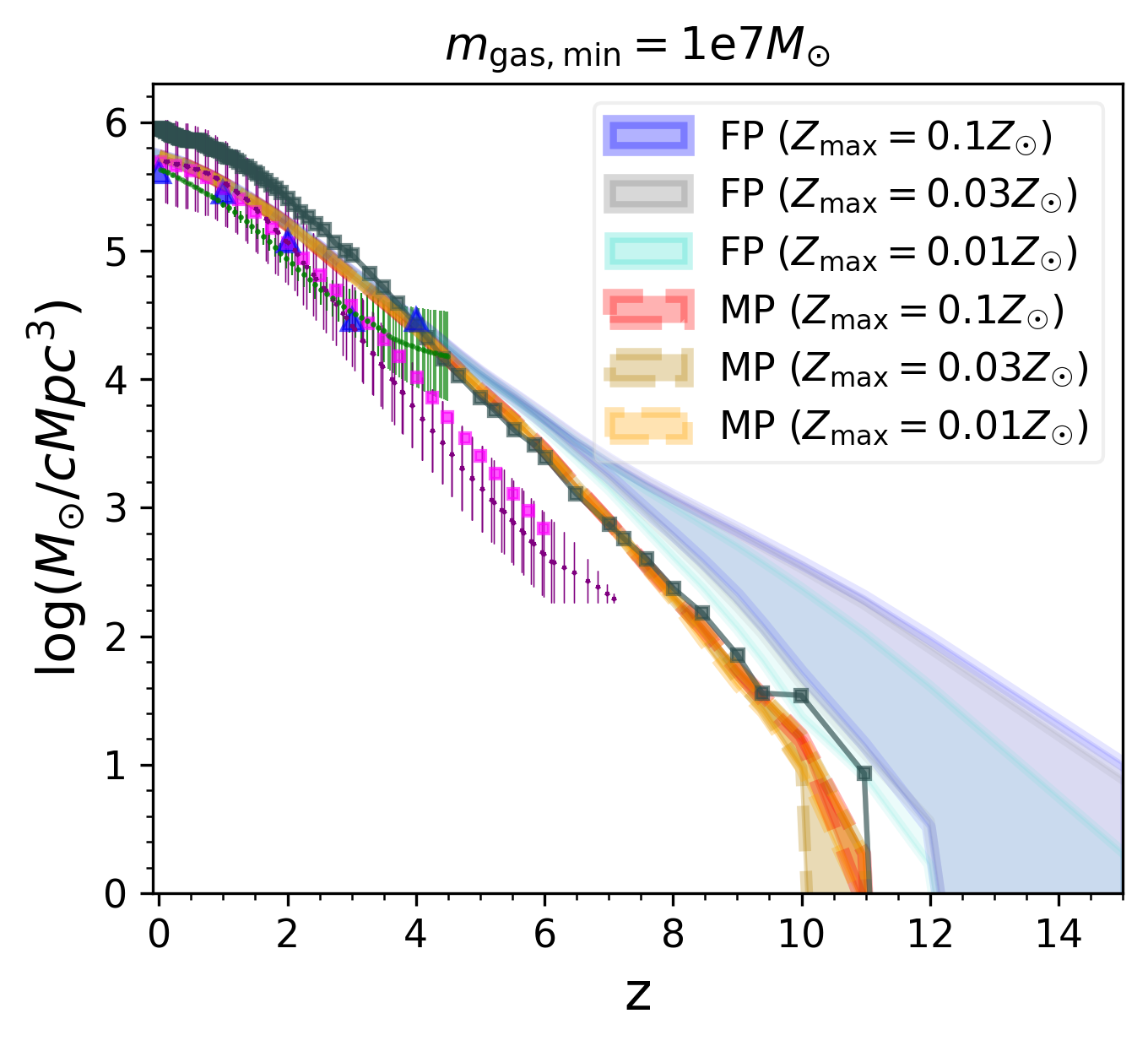}
\includegraphics[width=0.32\textwidth]{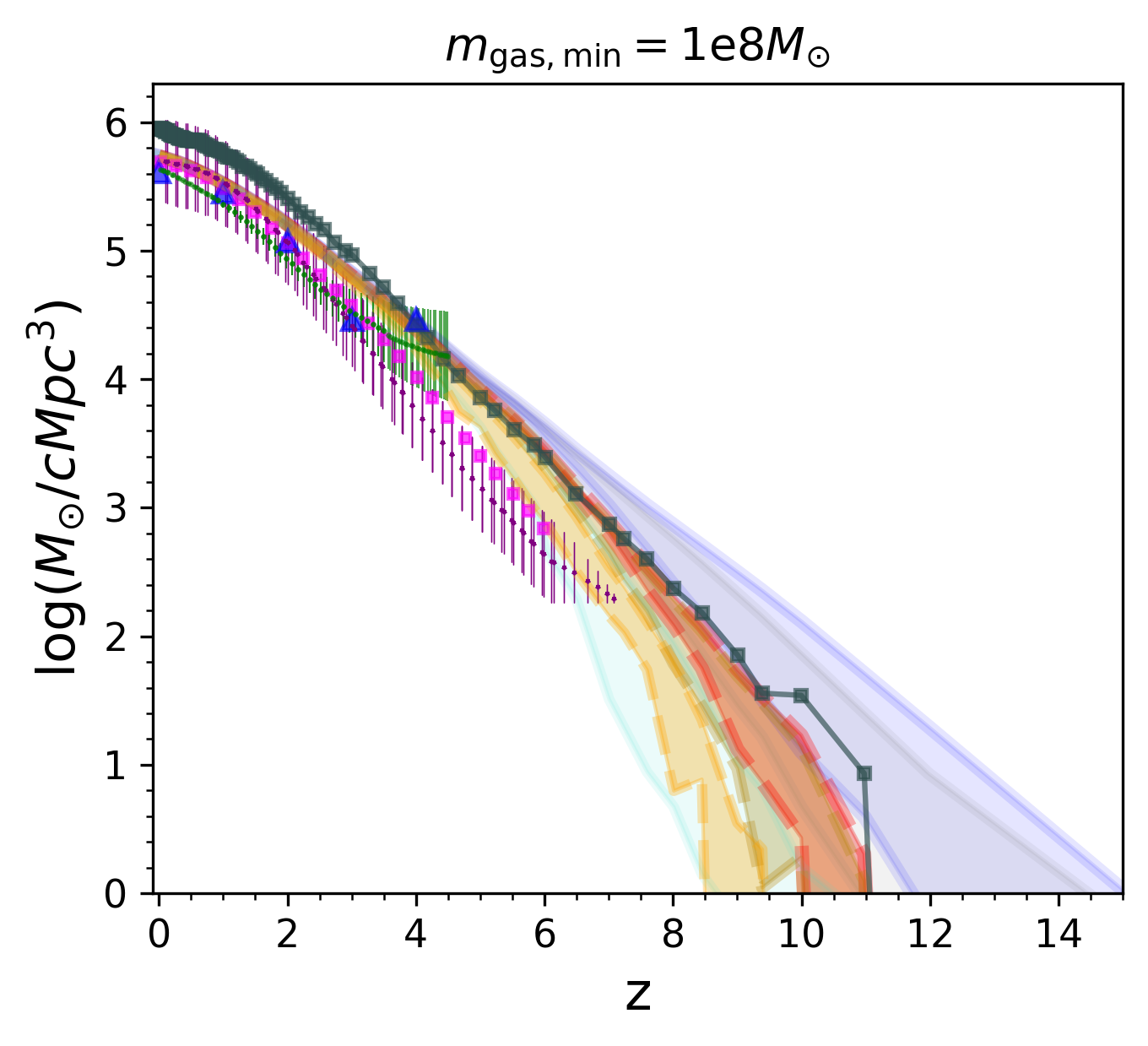}
\includegraphics[width=0.32\textwidth]{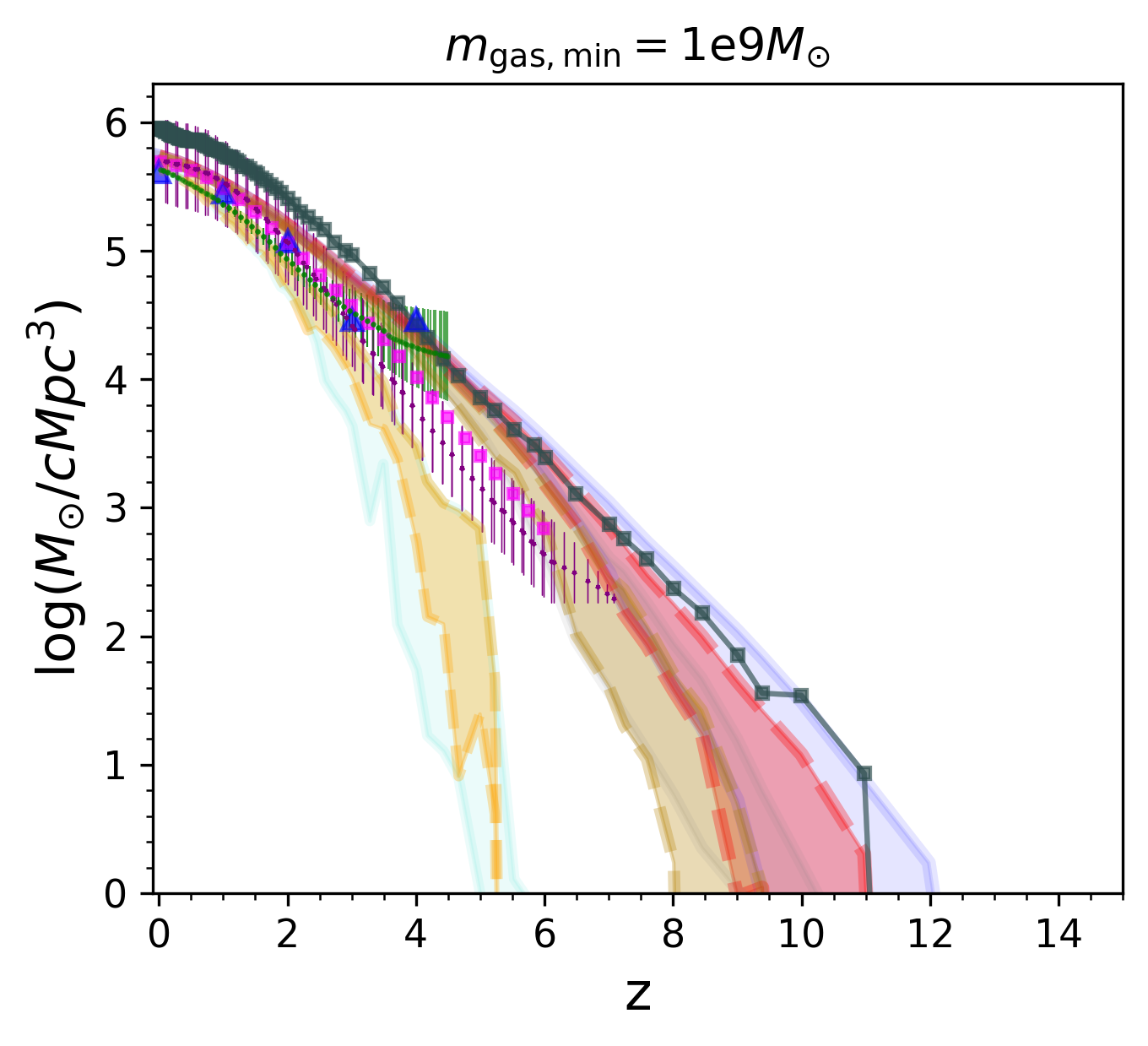}

\includegraphics[width=0.32\textwidth]{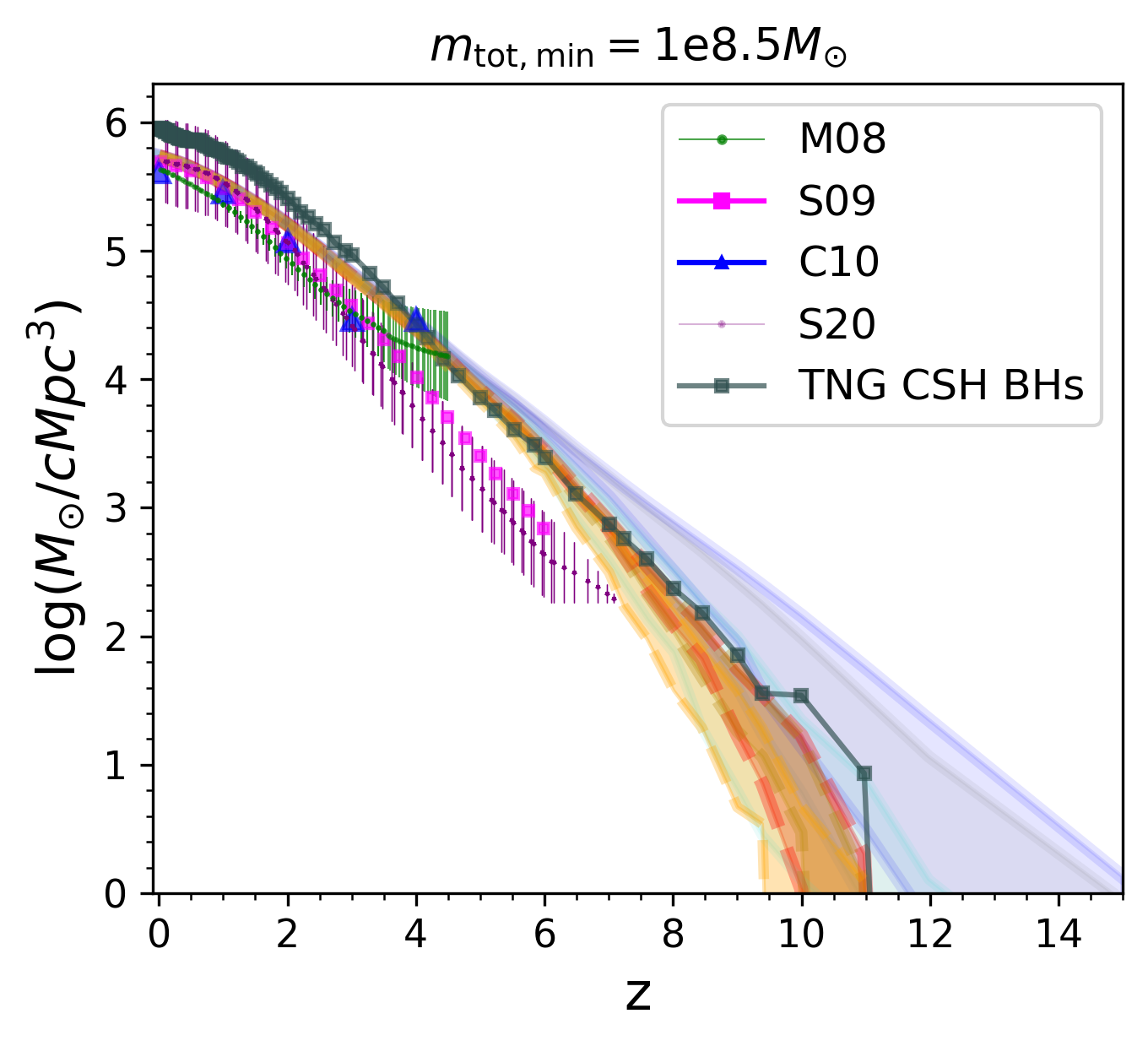}
\includegraphics[width=0.32\textwidth]{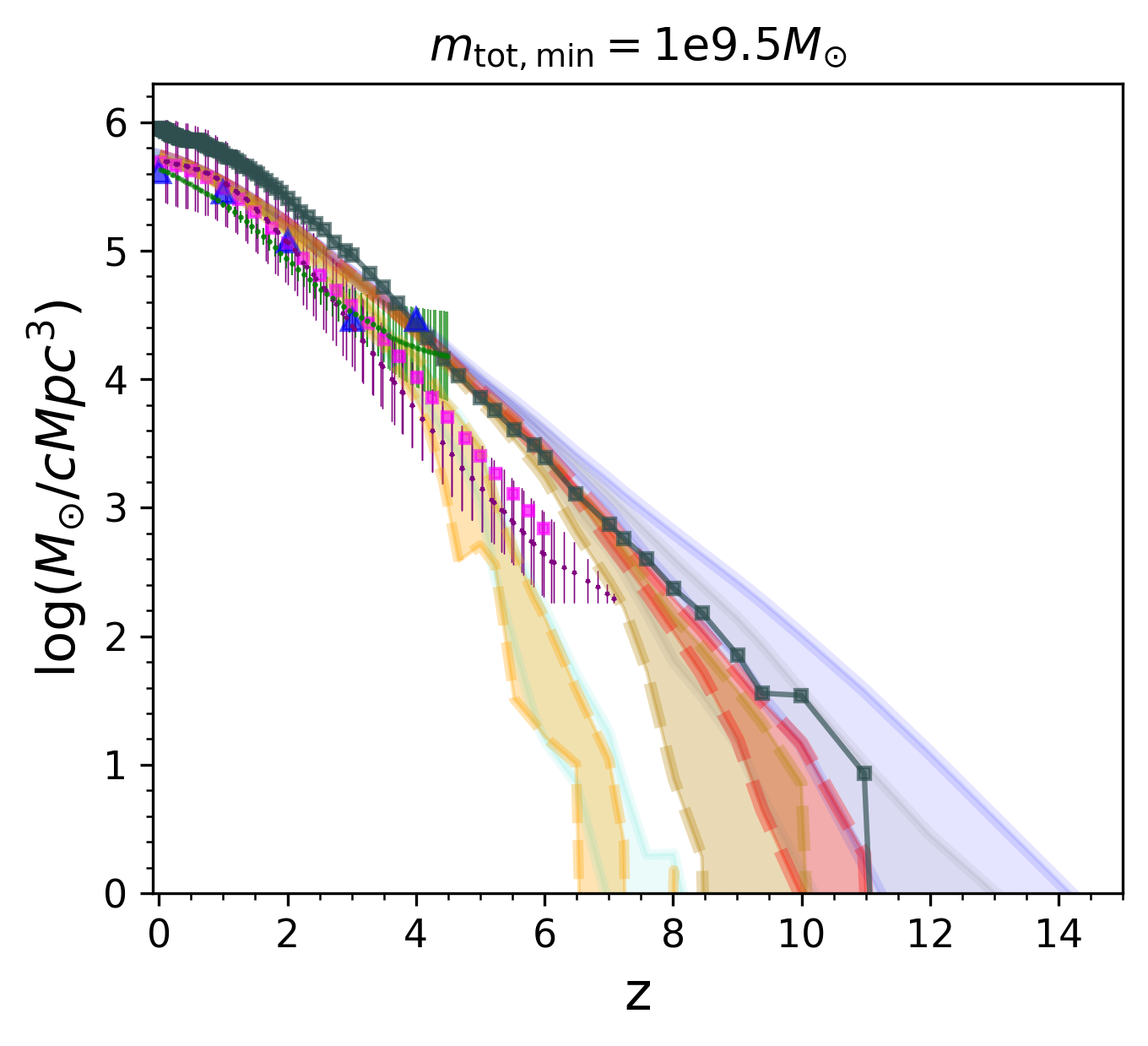}
\includegraphics[width=0.32\textwidth]{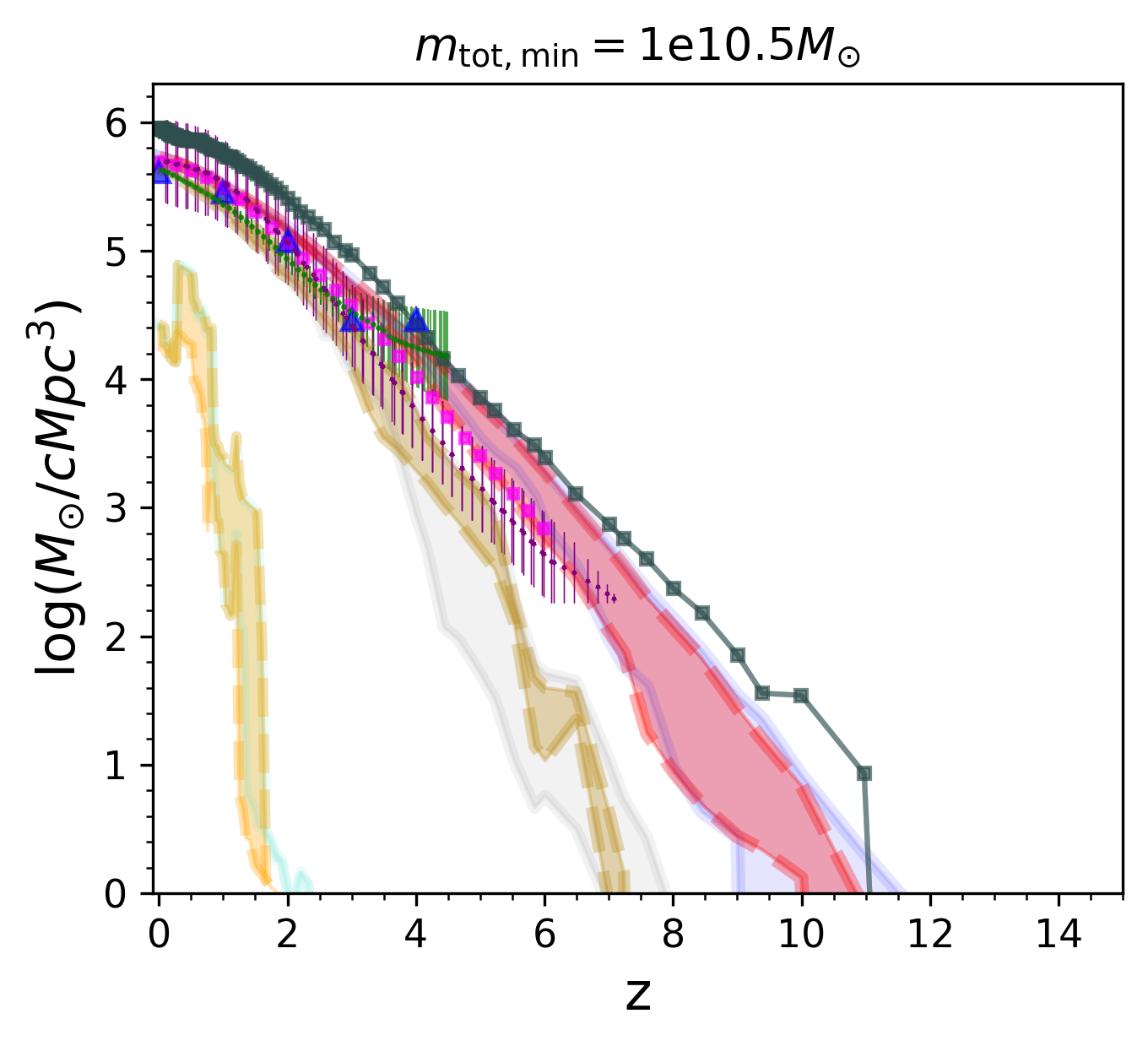}

\caption{Comoving mass densities are shown for models in the same style and corresponding to those in Figure~\ref{fig:number_densities.png}. All of the plausible seed models with consistently reasonable number densities also have 
mass densities consistent with  observations and TNG. Several model mass densities are in good agreement with TNG and empirical data. The large ratio of $<10^{5}$ \msun\ to $>10^5$ \msun\ BHs above $z\gtrsim 4$ in these models  emphasizes LISA as a key observational program for the predicted BH masses. Differently from the number density results, the empirical constraints on mass density are in much closer agreement with each other and with our seed models. Mass densities at $z\lesssim 4$ are also only slightly lower~(by a factor of $\sim1.5-2$) than the simulated TNG BHs. \texttt{mtot10.5\_Z0.01} is the only seed model which severely underpredicts the mass densities. It does not start producing seeds until $z\sim2$.
\label{fig:mass_densities.png}}
\end{figure*}

\begin{figure*}
\centering

\includegraphics[width=0.32\textwidth]{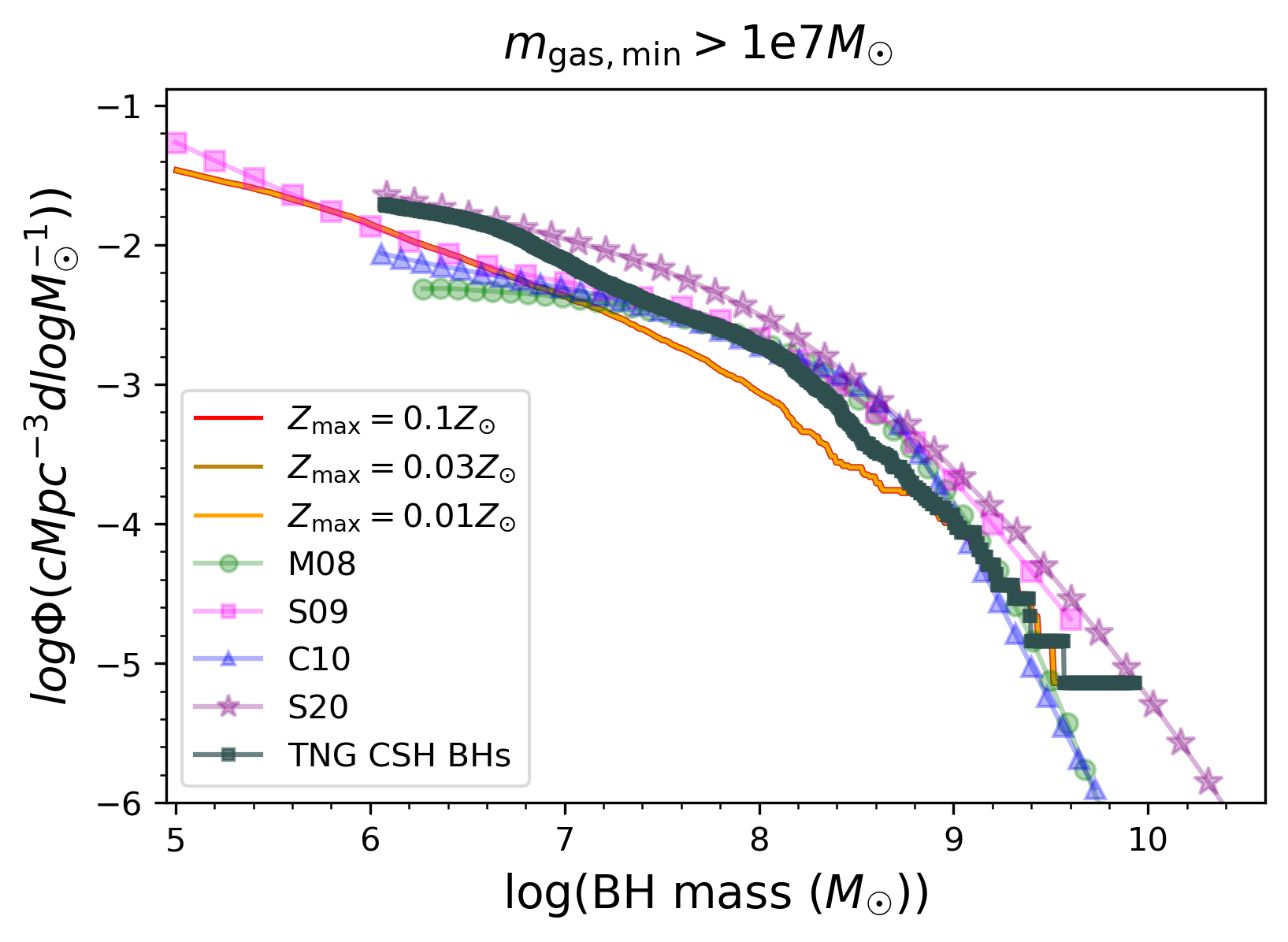}
\includegraphics[width=0.32\textwidth]{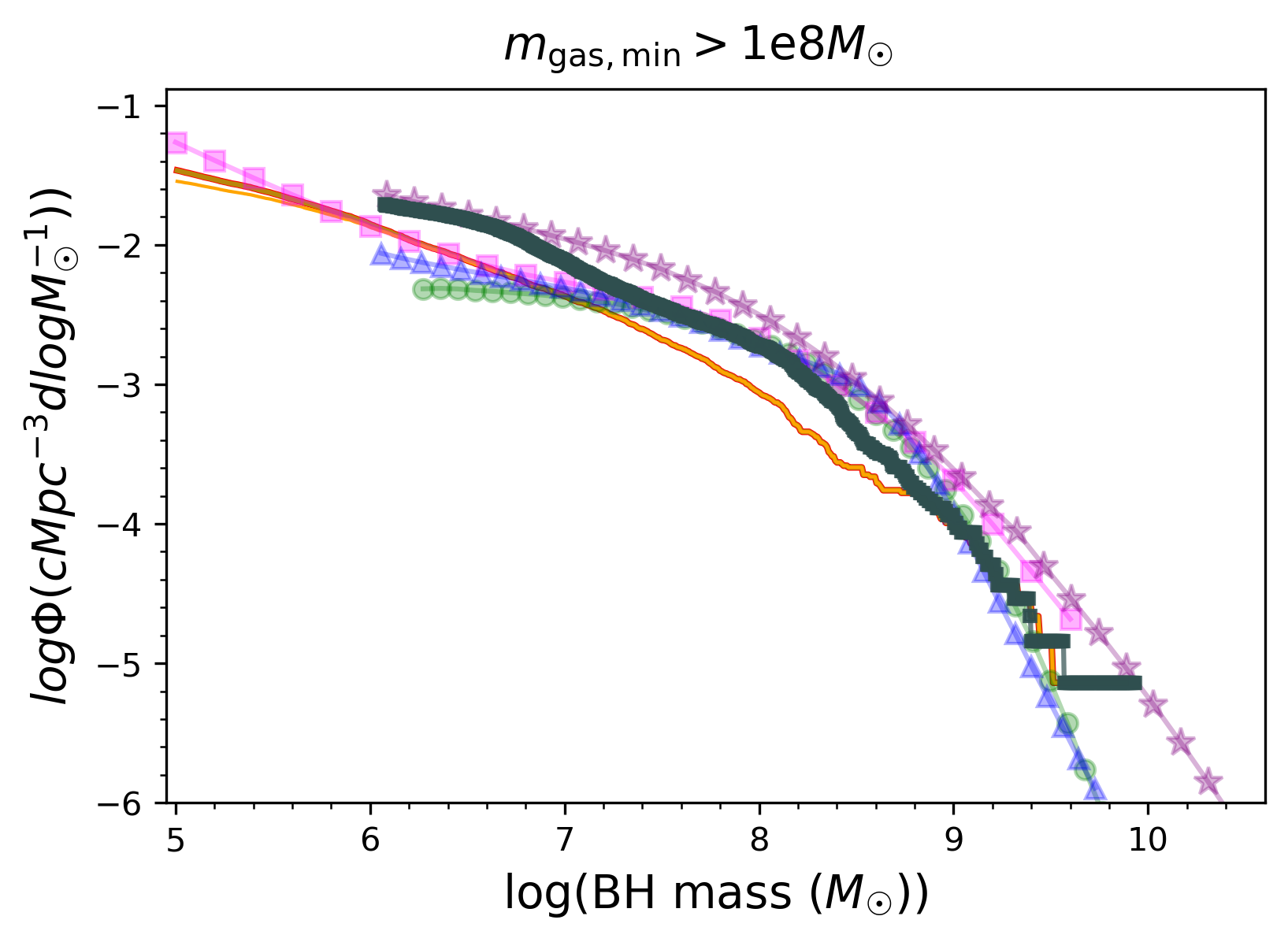}
\includegraphics[width=0.32\textwidth]{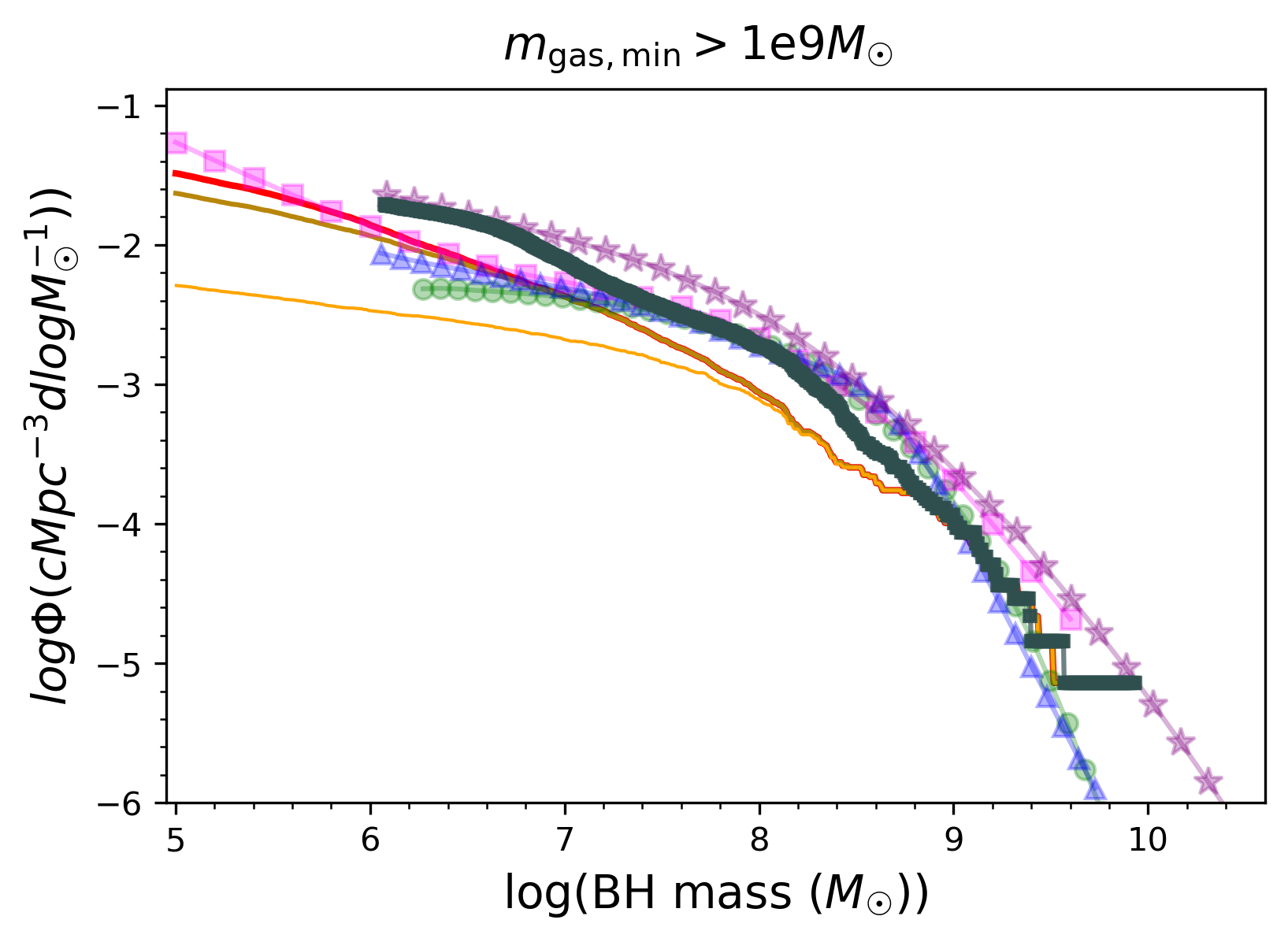}

\includegraphics[width=0.32\textwidth]{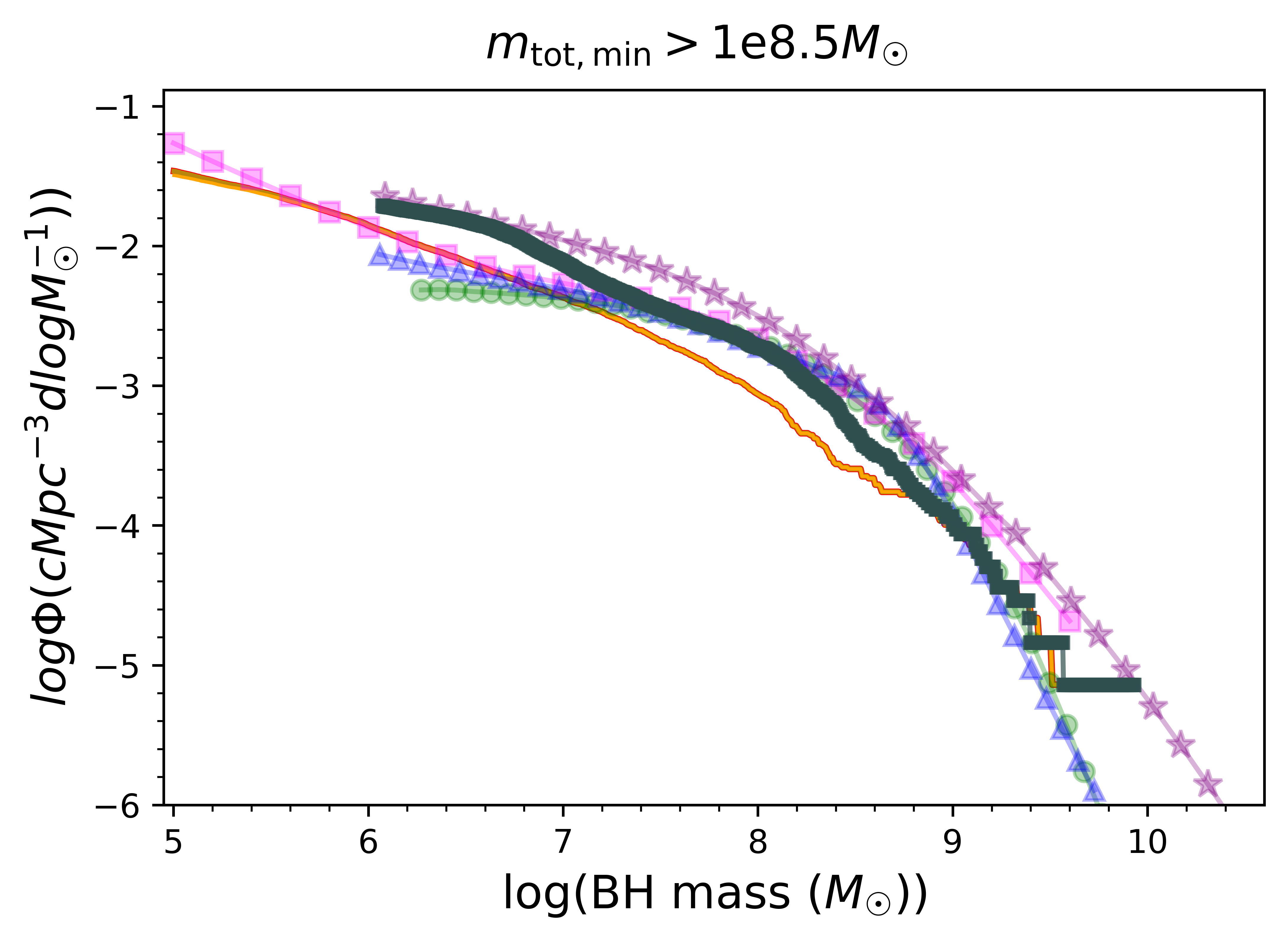}
\includegraphics[width=0.32\textwidth]{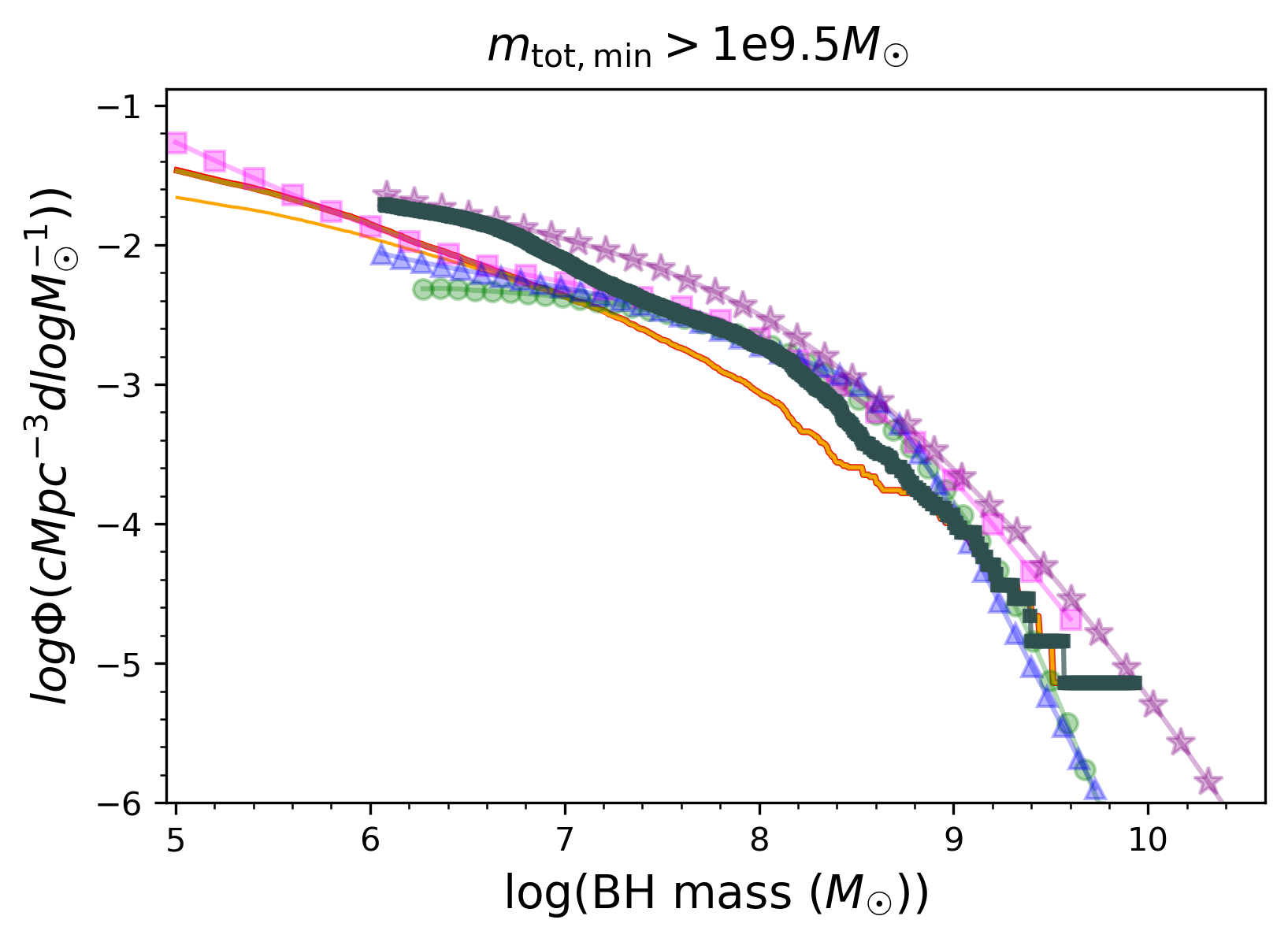}
\includegraphics[width=0.32\textwidth]{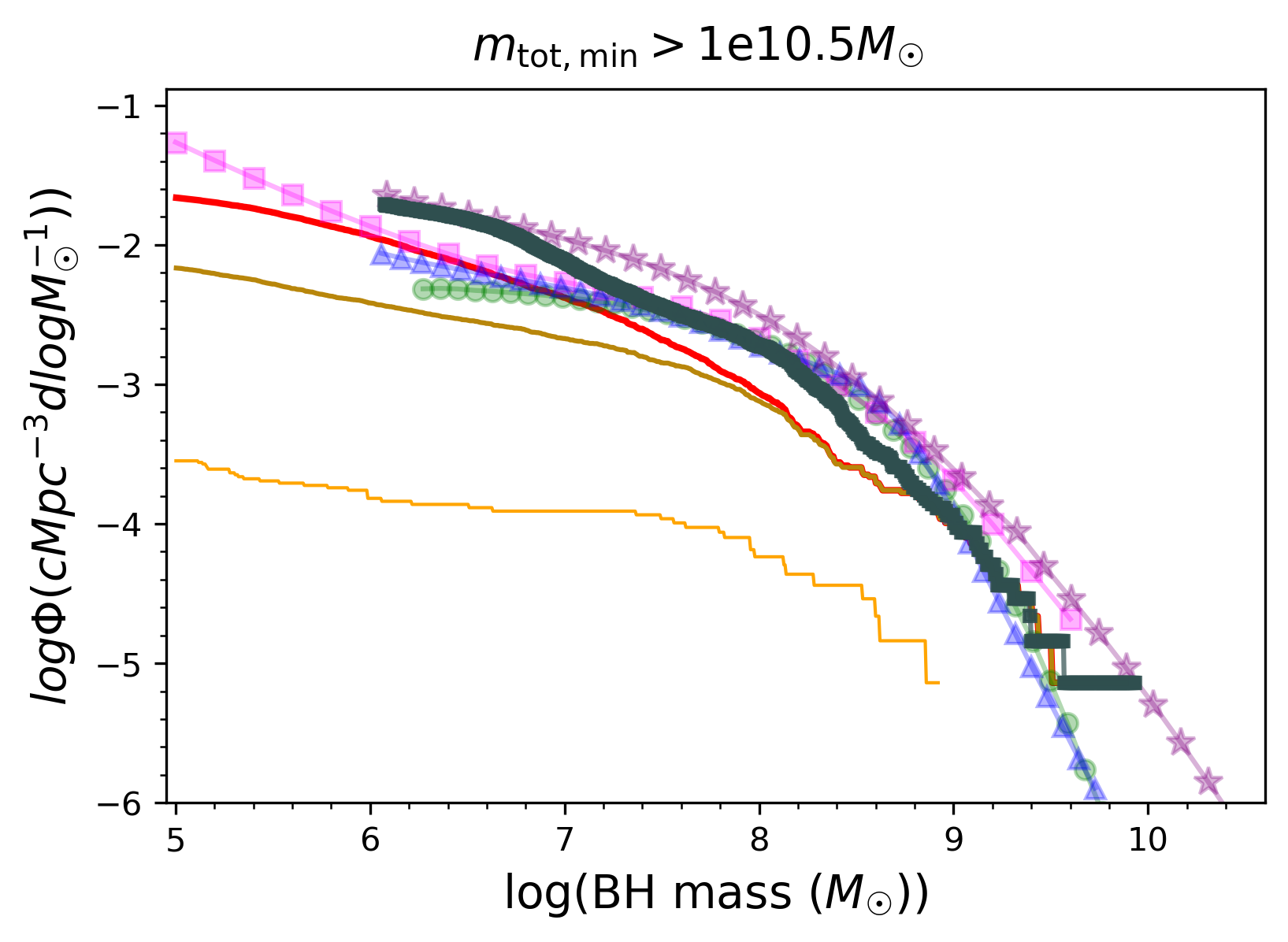}

\caption{In the same color scheme as Figure~\ref{fig:mass_densities.png}, the corresponding $z=0$ BHMFs are shown for the models. There are reasonable local BHMFs for all the plausible SAMs that are well within the empirical parameter space. At the most massive end~($\gtrsim10^9$ \msun), the model results agree well with TNG's. Below $\lesssim10^9$ \msun, the TNG BHMFs are slightly higher than those of the models ~(by factors of $\sim1.5-2$). At the most massive end~($\gtrsim10^9$ \msun), nearly all models are in best agreement with S20 with the exception of ones that produce too few BHs. $\zgas/$\zsun~$<10^{-2}$ hosts at $z=0$ are more rare and do not contribute to the most extreme mass end of the BHMF. BHMF results are as expected with the BH - stellar mass scaling relation model, which does not include scatter, and with the well-produced stellar mass function of TNG.
\label{fig:BHMFs.png}}
\end{figure*}

\subsection{Fiducial Suite of Semi-Analytic BH Seeding Models}
\label{ssec:fiducial}
Having validated our SAM by reproducing the TNG results, we are now finally ready to explore the wide range of physically motivated seed models from Section \ref{sec:Parameter_space}. In Figures ~\ref{fig:n_BH_intro.png}, ~\ref{fig:number_densities.png} and ~\ref{fig:mass_densities.png}, we analyze BH populations produced by these seed models in terms of their number density and mass density evolution. We consider two distinct types of BH populations:
\begin{itemize}\item The \textit{full population} of BHs formed in our SAMs, referred to as the ``FP BHs''. With all of the masses determined via the local BH scaling relations, the FP BHs have masses ranging from $\sim10$ \msun\ to $\sim10^{10}$ \msun. The lower BH mass limit is set by the adopted BH-stellar mass scaling relation and the requirement that the stellar mass be nonzero. 
\item BHs with masses $>10^{5}$ \msun, hereafter referred to as the \textit{massive population} of BHs or ``MP BHs.''
\end{itemize}
In the following subsections, we will systematically address the impact of our seed models on different aspects of the number density and mass density evolution of the resulting BH populations.

\subsubsection{Redshift evolution of BH number densities}

The shaded regions in Figure~\ref{fig:n_BH_intro.png} shows the number densities of FP~(cool colors) and MP~(warm colors) BHs predicted by our seed models. Different colors indicate models with different $Z_{\rm max}$ values, and the shaded region spans the range of stochastic seeding models from $f_{\rm seed} = 0.01$ to $1$. As expected, number densities increase quickly with time at the highest redshifts, as rapid halo growth drives the formation of new seeds. For most models, the 
number densities peak at redshifts between $\sim 2 - 7$, when halo enrichment slows the formation of new seeds, after which it decreases with time. This is due to a combination of several effects as identified in Section~\ref{sec:massmet_relations}: 1) seed formation is slowed by metal enrichment in halos, 2) star formation and feedback can reduce the amount of gas available to form seeds inside the halos, 3) the BHs undergo mergers with each other.

Generally, we see that the saturation in the BH number densities tends to happen at later times as the seeding criteria become more strict. This happens because of a combination of effects. First, the maximum number of halos~(with masses $>10^7$ \msun) available for seeding saturates at $z\sim6$~(revisit Figure~\ref{fig:H_frac_Z}, blue line). This essentially sets $z\sim6$ to be the ``saturation redshift" of the FP BH number densities for the most lenient seed models like \texttt{mgas7\_Z0.1} and \texttt{mtot8.5\_Z0.1}~(Figure~\ref{fig:n_BH_intro.png}, leftmost panels). But in stricter seed models, the BH occupation fraction in halos is lower at early times ($z\gtrsim6$), such that proportionally more halos are available to form new seeds at $z\lesssim6$. Therefore, for the stricter seed models like \texttt{mgas9\_Z0.1} and \texttt{mtot10.5\_Z0.1} (Figure~\ref{fig:n_BH_intro.png}, rightmost panels), the saturation in the FP BH number densities starts to occur at lower redshifts (i.e., $z\sim4$ and $z\sim2$, respectively). Irrespective of these trends, however, we see that more lenient seed models form more BHs at all redshifts.

\subsubsection{Impact of halo mass and gas mass seeding thresholds on the BH number densities}

Not surprisingly, the BH number densities tend to decrease with increasing halo mass~($\mtot$) and gas mass seeding thresholds~($\mgas$). The impact is generally stronger at higher redshifts, simply because the underlying halo mass functions are steeper. Additionally, the FP BH number densities are much more sensitive to the seeding criteria 
than the MP BH number densities. For example, as we go from the most lenient to the strictest $\mgas$, (i.e., $\mgas=10^7$ to $10^9$ \msun), the FP BH number densities can be suppressed by one to two orders of magnitude for $\zgas=0.1$~(blue shaded regions in Figure~\ref{fig:n_BH_intro.png}, top panels), whereas the corresponding MP BH number densities vary much less with $\mgas$ (red shaded regions in Figure~\ref{fig:n_BH_intro.png}, top panels). Overall, this is because increasing the mass threshold suppresses low-mass seed formation, which has a disproportionately stronger impact on the lower-mass FP BHs compared to the MP BHs. However, for the majority of the halo and gas mass thresholds, the BH population is dominated by the FP BHs. These are largely comprised of low-mass ($\sim 10-10^5$ \msun) BHs that are currently inaccessible to EM observations at high redshift. However, upcoming GW facilities like LISA will be sensitive to low-mass, high-redshift mergers, 
which will likely provide strong constraints on seed models. 

\subsubsection{Impact of gas metallicity threshold on the BH number densities}

We now compare the number density predictions of MP and FP BHs for two different gas metallicity thresholds for seeding, i.e., $\zgas=0.1~\&~0.01$. We can clearly see that when the halo and gas mass thresholds are increased, the metallicity threshold has a stronger impact on seeding. For example, among the varying-$\mgas$ 
models~(blue vs.~turquoise regions in the top panels of Figure~\ref{fig:n_BH_intro.png}), when $\mgas$ is $10^7$ \msun, decreasing $\zgas$ from 0.1 to 0.01 \zsun\ makes a very small difference in the number densities of FP BHs. For a higher $\mgas$ of $10^9$ \msun, $\zgas = 0.01$   \zsun\ produces up to $\sim1000$ times fewer BHs compared to $\zgas = 0.1$ \zsun. Overall, this is because more massive halos tend to be more metal enriched due to a more extensive history of star formation and evolution. As we can see in Figure~\ref{fig:massmet}, the vast majority of $>10^7$ \msun\ halos have metallicities $<0.01$ \zsun. In contrast, a very small minority of $>10^{10}$ \msun\ halos have metallicities $<0.01$ \zsun. 

The impact of the metallicity criterion substantially decreases with time in general. In fact, for the lowest $\mtot$ and $\mgas$, models with different $\zgas$ produce similar results at $z\sim0$. This trend is not surprising because even though more BHs are formed at earlier times in models with higher $\zgas$, cosmic evolution causes them to merge with each other as their host halos merge. As a result, the differences in the  high-$z$ number densities seen for models with different $\zgas$, washes out over time. To that end, note that our models assume prompt mergers amongst BHs within the same halo, thereby excluding wandering off-center BHs, or BHs in satellite galaxies. If these populations were included, we could expect the impact of $\zgas$ to persist more strongly at lower $z$.  

\subsubsection{Impact of seed probability on the number densities}

Here we examine the impact of probabilistic seeding ($f_{\rm seed}$) on the BH number densities. Note that the seed probability is applied~(as a random draw) on every descendant along a given tree. In the absence of a metallicity criterion for seeding, applying such a probabilistic seed criterion would simply lead to an effective delay in the seed formation along a tree branch. However, the presence of a metallicity criterion dictates that the formation of a seed on a tree branch hinges upon the rate of metal enrichment along that branch. If the tree branch undergoes rapid metal enrichment and the seed probability is low enough such that the branch is already enriched with metals by the time sufficient random draws are available to place a seed, then no seed will form on that particular branch at all. 
BH number density predictions for seed probabilities of 1 and 0.01 are shown as the upper and lower limits of the shaded regions in Figure~\ref{fig:n_BH_intro.png}. We can see that the shaded regions tend to shrink as redshift decreases; in fact, by $z\sim0$, both seed probabilities produce very similar number densities. This suggests that for a seed probability of 0.01, metal enrichment does not occur rapidly enough to completely prevent seeding on the vast majority of the tree branches. 

It is useful to compare the impact of seed probability vs that of gas metallicity threshold since the former is intended to account for additional physics~(halo growth, star formation and metal enrichment) that can influence seeding. Notably, we find that the impact of reducing $\zgas$ from $0.1$ and $0.01$ on the number densities is stronger than that of reducing $f_{\rm seed}$ from 1 to 0.01. Nevertheless, both parameters do have a significant impact, particularly at the highest redshifts. This motivates the need for exploring the variety of other physics that can impact BH seeding such as UV radiation, gas angular momentum, dynamical heating, etc.; this will be focus of future work. Note that because $f_{\rm seed}$ is the seeding probability applied at each snapshot, it implicitly depends on the time resolution of TNG snapshots. In other words, a given value of $f_{\rm seed}$ would not have the same physical meaning in a simulation with higher or lower time resolution.

\subsubsection{Comparison of the number density predictions to TNG and empirical data}

We finally compare the BH number densities predicted by our seed models to empirical data shown in Figure~\ref{fig:n_BH_intro.png}. Note that observations have thus far not been able to probe BH populations $\lesssim 10^5$ \msun. Therefore, it is not surprising that for most of our seed models, the number densities of FP BHs substantially exceed that of the empirical data. To that end, the MP BHs offer a fairer comparison to the empirical data. However, recall from Section~\ref{tng_vs_model} that even amongst the empirical data, the various published measurements vary by factors of up to $\sim4$ at $z\sim0$. 
Hence, the following serves merely as a broad comparison between simulations and observations.

In Figure~\ref{fig:number_densities.png}, we replot the predicted BH number densities already shown in Figure~\ref{fig:n_BH_intro.png}, but here we solely focus on the MP BHs. We also include an additional, intermediate model with $\zgas=0.03$ \zsun. The most lenient models like \texttt{mgas7\_Z0.1} and \texttt{mtot8.5\_Z0.1} predict BH number densities that differ from empirical constraints at $z\sim0$ by factors of up to $\sim3$. 
Note that only S09 attempt to include $10^5-10^6$ \msun\ BHs in their analysis; the others use $10^6$ \msun\ as the lower limit on BH mass. At higher redshifts, the empirical constraints become even more uncertain. 
This undoubtedly contributes to the fact that 
a substantial majority of our models predict higher BH number densities than are obtained from empirical constraints, with the exception of S09 at $z\sim0$. This includes the \texttt{mgas9\_Z0.03} and \texttt{mtot10.5\_Z0.1} models for which the number density evolution most closely resembles that from TNG. 

In fact, only two of our~(strictest) models predict BH number densities within the range spanned by the different empirical constraints at intermediate redshifts of $z\sim 1 - 4$; these are \texttt{mgas9\_Z0.01} and \texttt{mtot10.5\_Z0.03}~(top right and bottom right panels,  respectively, in Figure~\ref{fig:number_densities.png}). 
However, these models underpredict the $z=0$ number densities, and they do not begin forming BHs until $z<8$, which is inconsistent with recent discoveries of very high-redshift AGN with JWST \citep{2023arXiv230308918L,2023ApJ...942L..17O,2023arXiv230609142S,2022Natur.604..261F,2023MNRAS.519.4753T}. This underscores the importance of low-mass BHs in calculations of the BH number density. This is even more true as we go to higher redshifts, where the empirical constraints become increasingly uncertain. Nevertheless, we can still fully rule out one of our strictest models (i.e., \texttt{mtot10.5\_Z0.01}), which predicts number densities substantially below all of the empirical constraints at all redshifts; this is because the seed production does not start until $z\sim2$~(bottom right panels of Figures \ref{fig:n_BH_intro.png} and \ref{fig:number_densities.png}). 

Overall, the foregoing results demonstrate that BH number densities are sensitive to different 
seeding scenarios, particularly at higher redshifts wherein the variations amongst our seed models  are large (exceeding $\sim100$ and $\sim10$ for FP and MP BHs, respectively). Continued observations with JWST are expected to reduce these uncertainties. Additionally, observations of even lower mass~($\sim10^4-10^6$ \msun) BHs with upcoming LISA and proposed EM facilities such as Lynx can pose even more stringent constraints on our seed models. At the other end, our predicted number densities may also be impacted by the modelling of physical processes such as star formation, metal enrichment and stellar feedback. In future work, we will continue to use our newly built framework to systematically explore the impact of all these processes. This will be crucial preparation for the wealth of observational data that we expect from the coming decades.

\subsubsection{Mass density evolution}

Mass density evolution  (Figure~\ref{fig:mass_densities.png}) varies much less between the seed models, compared to their number density evolution. 
This is especially true at $z\lesssim 4$ for both MP and FP BHs. 
converge to similar values at $z\lesssim 4$. 
Even in seed models with the most lenient total and gas mass cuts, for which number densities are dominated by BHs $<10^5$ \msun, we still see similar $z=0$ mass densities between the MP and FP BHs. This implies that for all of our seed models, the mass densities at $z\lesssim 4$ are dominated by BHs significantly more massive than $10^{5}$ \msun. Notably, the empirical constraints on mass density (most of which extend to $z\lesssim4$) are also in much closer agreement with each other and with our seed models, compared to the number densities. 
These mass densities are also only slightly lower~(by a factor of $\sim1.5-2$) than the simulated TNG BHs. The only seed model that severely underpredicts the mass densities is \texttt{mtot10.5\_Z0.01}, since it does not start producing seeds until $z\sim2$. 

Recall here that the BH masses are a fixed fraction of the host stellar masses based on the local $M_*-M_{bh}$ scaling relations. Therefore the agreement between the seed models and the empirical measurements is not surprising, given that the underlying TNG galaxy formation model successfully reproduces the observed stellar mass functions and the cosmic star formation rate densities at $z\lesssim4$ \citep{2018MNRAS.474.3976G}.

At $z\gtrsim4$, we see more variations in the mass density predictions between the different seed models. In this regime, the only available empirical constraints are from S20 that span from $z\sim0-7$. We see that seed models with the most lenient mass cuts~($\mtot=10^7~M_{\odot}$ and $\mgas=10^{8.5}~M_{\odot}$) somewhat overestimate the mass densities compared to S20. But the more restrictive seed models do produce reasonable agreement with S20. Notably, only a few seed model predictions fall within the error bars of S20 over the entire redshift range covered by their measurements. Future constraints on the mass densities at high redshifts using facilities like JWST will help us better discriminate between the different seeding scenarios. Finally, we also note that at $z\gtrsim4$, the mass densities for the FP BHs are significantly higher than the MP BHs. This implies that at these redshifts, the overall mass densities are largely contributed by low-mass ($<10^5$ \msun) BHs, which will be difficult to access with EM observatories. LISA observations are therefore going to play an essential role in constraining the mass densities at these redshifts.

\subsection{Local BHMFs produced by Fiducial Suite of BH Seed Models}
\label{ssec:bhmf}

The $z=0$ BHMFs for our models are shown in Figure~\ref{fig:BHMFs.png} with the same color scheme as Figure~\ref{fig:mass_densities.png}. The probabilistic seed models are omitted since they produce the same results as the non-probabilistic models by $z=0$.  There is very little variation in the predicted BHMFs among most of our seed models. Overall, our seed model BHMFs are in broad agreement with the empirical BHMFs. As we make the seed models more restrictive, we start to see underpredictions of the BHMF, starting at the lowest mass end~(lower middle panel). For the strictest seed models~(right panels), we see appreciable variations in the BHMF over a larger range of BH masses. Not surprisingly, there is generally more variation at the low mass end due to greater retention of the memory of the initial seed mass. Our seed models predict similar BHMFs as the TNG model at the most massive end~($\gtrsim10^9$ \msun). 
At the most massive end~($\gtrsim10^9$ \msun), both TNG and nearly all of our seed models lie within the range of the empirical BHMFs; this is with the obvious exception of \texttt{mtot10.5\_Z0.01} that produces too few seeds for any significant BH population to form. 
All of our seeding models do, however, produce a slightly shallower "knee" in the BHMF relative to observations and TNG, which may be a result of the direct scaling between BH and stellar mass in our models.

In any case, it is fair to say that given the spread within the empirical BHMFs themselves, most of our seed models do a reasonable job in reproducing the local BH population. Similar to the mass density evolution, the above results are also not surprising given that 1) the BH masses are assigned to be a fixed fraction of the stellar mass consistent with the local $M_*-M_{\mathrm{bh}}$ relations, 2) TNG stellar mass functions are consistent with the observational constraints. 
Recall that we have imposed a simple zero-scatter, $z=0$ $M_{\rm BH}-M_*$ scaling relation to populate BHs in the galaxy merger tree.

\section{Conclusions}
\label{sec:conclude}

In this work, we build novel semi-analytic BH seed models that form BHs and trace their evolution along galaxy merger trees within the TNG50 volume of the IllustrisTNG simulation suite. We systematically explore a wide range of criteria for seeding a BH in TNG halos. We consider models that seed a BH in each halo that exceeds minimum thresholds in gas mass ($\mgas=10^7-10^9$ \msun) and total mass ($\mtot=10^{8.5}-10^{10.5}$ \msun), with gas metallicities less than a maximum limit ($\zgas=0.1,0.03,0.01$ \zsun). We treat the BHs in our models independently from those in TNG, and we also make the simplifying assumption that at most one BH is present in each halo (i.e., we consider only the total BH mass per halo). The models are motivated by the expectation that popular theoretical seeding channels such as Pop III, NSC and DCBH seeds form in halos with low metallicity and dense (star-forming) gas. The halo mass cuts ensure that the seeding takes place regions with deep enough gravitational potentials and that no seeds form in spuriously identified gas clumps outside of dark matter halos. 
The gas mass cuts ensure that there is sufficient gas in the halo, a small fraction of which is presumed to actually form the BH seed. To account for the possibility that additional criteria may be required to form BH seeds, we also consider models in which each halo that meets all other criteria forms a BH seed with probability $f_{\rm seed} = 0.01$.  Lastly, we also ensure that the seeded halos have at least one star particle, to ensure that these halos have a prior history of assembling dense star forming gas, and because we assign BH masses based on a simple scaling with the host stellar mass.

We first validated our approach by using the original TNG50 seeding criterion in our semi-analytic framework (i.e., seeding BHs in $m_{\rm tot}>5\times10^{10}$ \msun $h^{-1}$ halos). 
When these BHs are populated in our halo merger trees, we find that the resulting BH counts are consistent with the BH population produced in the original TNG50 run to within $4$\% at $z=0$.
We then proceed to make predictions of BH populations for a wide range of seed models and compare them to empirical constraints from AGN observations (M08, S09, C10, and S20). 

Here we highlight our main conclusions:

\begin{itemize}

\item A wide range of seeding criteria produce number densities of massive BHs ($> 10^5$ \msun) that are broadly comparable to current empirical measurements. Only one of our strictest models
(\texttt{mtot10.5\_Z0.01}) completely fails to produce enough BHs at any epoch. The most lenient models produce somewhat more BHs than the TNG simulations as well as empirical measurements at $z\sim0$, with the exception of S09. However, note that there is uncertainty among the empirical measurements at $z\sim0$, with very few constraints at the low-mass end ($\sim 10^5 - 10^6$ \msun~, which S09 includes). At higher redshifts, the empirical constraints are even more uncertain. Most of our models predict higher number densities than these measurements, especially at high redshift. This tension reflects the large population of low-mass BHs in our models, and the dearth of empirical data on this population.
    
\item Just as the massive BH populations in our models are dominated by BHs at the low-mass end ($\sim 10^5 - 10^6$ \msun), when we consider the full population of BHs in our model (down to $\sim 10$ \msun), we find that 
the BH number densities are dominated by low-mass ($\sim 10-10^5$ \msun) BHs. This low-mass population is also more sensitive to changes in the halo or gas mass seeding thresholds. These $< 10^5$ \msun\ BHs would be difficult to detect with EM observations, but mergers between them would in many cases be observable with LISA, LIGO-Virgo-KAGRA, and next-generation ground-based GW detectors. We will quantify massive BH merger rates for our models in forthcoming  work.

\item Much less variation is seen in the BH mass densities, all of which converge to a narrow range of values at $0\lesssim z \lesssim 4$ consistent with empirical estimates (this excludes the aforementioned strictest \texttt{mtot10.5\_Z0.01} seed model). The good agreement in mass densities is a natural consequence of our BH mass growth model in which BH masses simply trace the host stellar mass, given the success of TNG simulations in reproducing the observational constraints for the galaxy stellar mass function and cosmic star formation rate density. However, at higher redshifts~($z\gtrsim4$), our seed models start to diverge in their mass density predictions for the massive $>10^5$ \msun\ BHs ~(up to nearly 2 orders in magnitude). At these redshifts, it is the low mass BHs that dominate the BH mass density, particularly for the more lenient seed models. This underscores the importance of LISA for the potential detection of these low mass BHs to constrain the high-$z$ BH mass density, and hence the underlying seeding channels.

\item Our BHMFs are very similar to the TNG BHMFs at the high-mass end ($\gtrsim 10^9$ \msun), but our model BHMFs are consistently lower than those in TNG at lower masses. This is also reflected in the slightly lower BH mass densities relative to TNG, which seeds only massive BHs ($8 \times 10^5$ \msun \hinv). Both our BHMFs and the TNG BHMF fall within the range of empirical measurements for the majority of our seed models. Again, these comments exclude the strictest \texttt{mtot10.5\_Z0.01} seed model that produces too few seeds. Additionally, the \texttt{mtot10.5\_Z0.03} and \texttt{mgas9\_Z0.01} models also somewhat under-produce the $\lesssim 10^8$ \msun\ BHs. At the other end, the more lenient seeding models produce nearly identical $z=0$ BHMFs, which reflects their consistent $z=0$ BH occupation fraction of essentially unity for halos resolved in TNG.

\item Considering together the BH mass and number densities and BHMFs, we still find that a wide range of seeding models produce BH populations in reasonable agreement with observations. In most cases, our models do produce higher BH number densities than those inferred from empirical data, but this is significantly influenced by the inclusion of $10^5$ - $10^6$ \msun\ BHs in our massive BH population, a mass regime where few empirical constraints exist.
We note that a combination of the varying-$\mgas$ and  varying-$\mtot$ 
cuts produces similar results to those presented here. In nearly all cases, reasonable $z=0$ BH populations are produced when combining these mass cuts with a maximum gas metallicity ranging from $0.01 - 0.1$ \zsun\  and a seeding probability from $0.01 - 1$. The exception is the strictest metallicity cut ($\zgas < 10^{-2}$ \zsun) combined with the strictest mass cuts ($m_{\rm tot} > 10^{10.5}$ \msun\ or $\mgas = 10^{9}$ \msun); these models produce few if any BHs at $z>6$ and cannot reproduce the $z=0$ BH population. 

\end{itemize}

Until the BHMF and its redshift evolution are more well-determined, this uncertainty will continue to be a barrier for models of BH formation and evolution, particularly in the low-mass and high-redshift regimes. JWST is pushing the envelope, being able to observe both bright and faint quasars earlier than previously possible \citep{2023arXiv230308918L,2023ApJ...942L..17O,2023arXiv230609142S,2022Natur.604..261F,2023MNRAS.519.4753T}.
 Paired with GW observations of SMBH binaries expected from LISA as far back $z\approx20$, this will greatly increase our understanding of BH populations at
early cosmic times. In turn, these data will constrain theoretical models of BH formation and early evolution, allowing us to probe the elusive origins of massive BHs. 

\section*{Acknowledgements}

AE, LB, \& AB acknowledge support from NSF award AST-1909933, and LB acknowledges support from the Research Corporation for Science Advancement under Cottrell Scholar Award \#27553. We also thank Paul Torrey and Luke Kelley for helpful discussions on the results.

\section*{Data Availability}
Data from the sublink and merger tree catalogs for the TNG50 simulation used in this project may be found on the TNG Project website: \href{https://www.tng-project.org/data/}{https://www.tng-project.org/data/}.

Scripts that retrieve descendants from the TNG merger trees may be found in this Github repository: \href{https://github.com/akbhowmi/arepo$\_$package}{https://github.com/akbhowmi/arepo$\_$package}. 
 
\bibliography{bibliography}

\appendix

\section{Seeding in (Central) Subhalos versus Halos}
\label{appendix:csh}

\begin{figure*}
\centering
\includegraphics[width=16 cm]{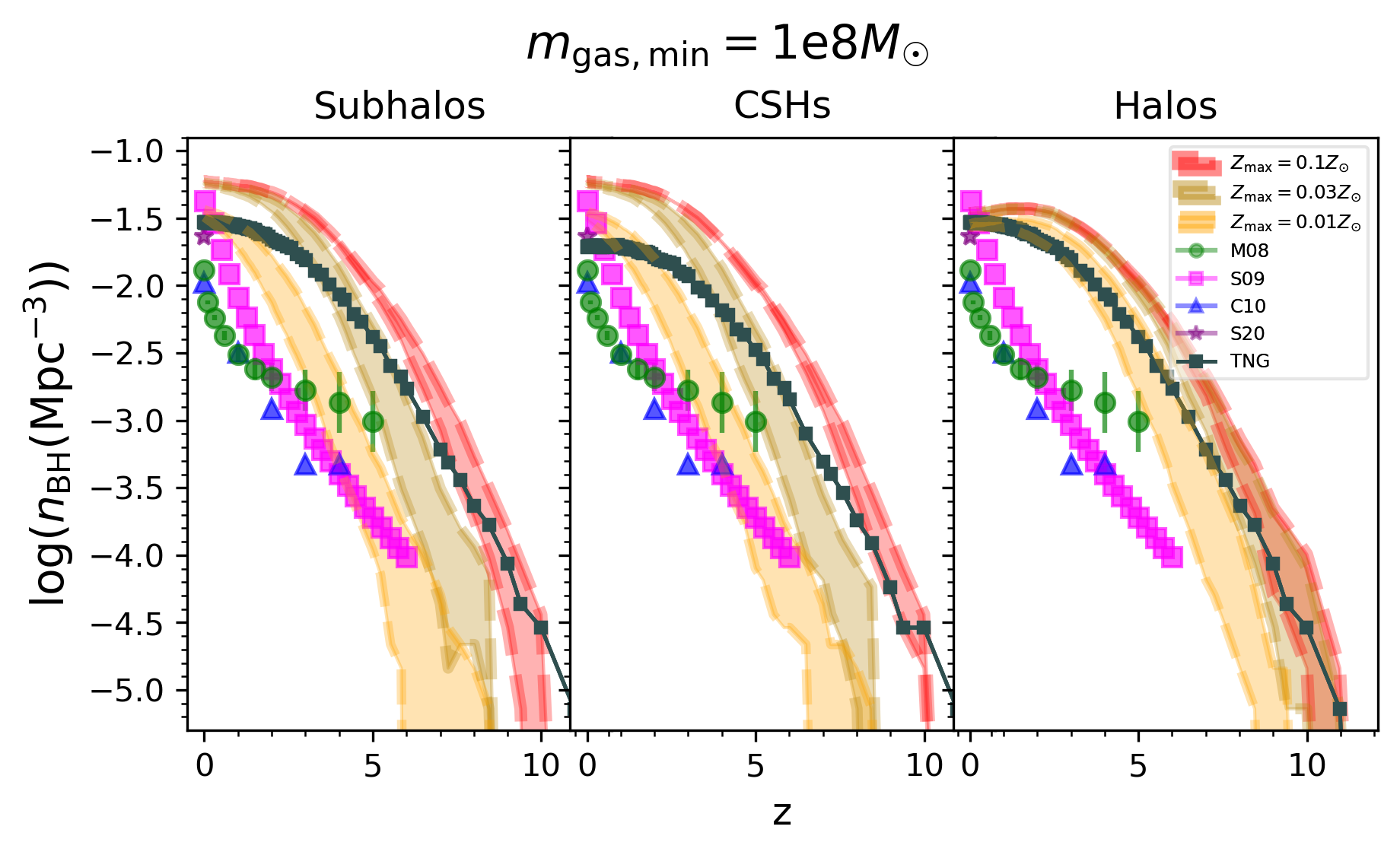}
\caption{ Number density evolution is shown for different hosts in the models where $m_{\rm{gas,min} }=10^{8}$ \msun. There is good agreement between the different host models, partly because of the model assumption of one BH per host. The halo models seed BHs relatively earlier than the other hosts and consistently agree best with TNG and S20 results. These model results highlight the reduction/enhancement of seeding sites between the different hosts. For example, this is naturally a more lenient gas mass cut for halos compared to CSHs, which explains the earlier BH seeding. The halo number densities are lower than in the CSHs at later times, since the group metallicity average is across all subhalo gas cells that belong to that particular halo or galaxy group. There is also less discrepancy between the halo models with different host metallicity criteria, compared to the same corresponding subhalo and CSH models.
\label{fig:host_comparison}}
\end{figure*}

In Figure~\ref{fig:host_comparison}, we reproduce one of the best-matching lenient mass cut models with respect to TNG and empirical results, except now across different hosts halos/subhalos/CSHs. 
The MP is shown for SAMs where $m_{\rm{gas,min} }=10^{8}$ \msun. The results are robust across all the halo/subhalo/CSH model hosts that seed BHs early enough and cross the observational parameter space. This is partly because we restrict seeds to one per host. The TNG halo number densities have less discrepancy between models with different host metallicity criteria, compared to the CSHs and subhalos.
Generally, the halo models produce more BHs and form them earlier than the other hosts. The halo models produce lower BH number densities at later times, which specifically varies between seed models, but happens around $z\sim3$ in this case. This could be due to halo model's capacity for lower averaged metallicities when metallicities vary largely between the galaxies in the galaxy groups. By contrast, the subhalo group metallicities are more simplistically averaged out to the max radius of that particular subhalo. These differences in how the metallicities are averaged produce an apparent difference in the larger variations between metallicity criteria results.

\section{Resolution Convergence}
\label{sec:res_converge}

\begin{figure*}
\centering
\includegraphics[width=\columnwidth]{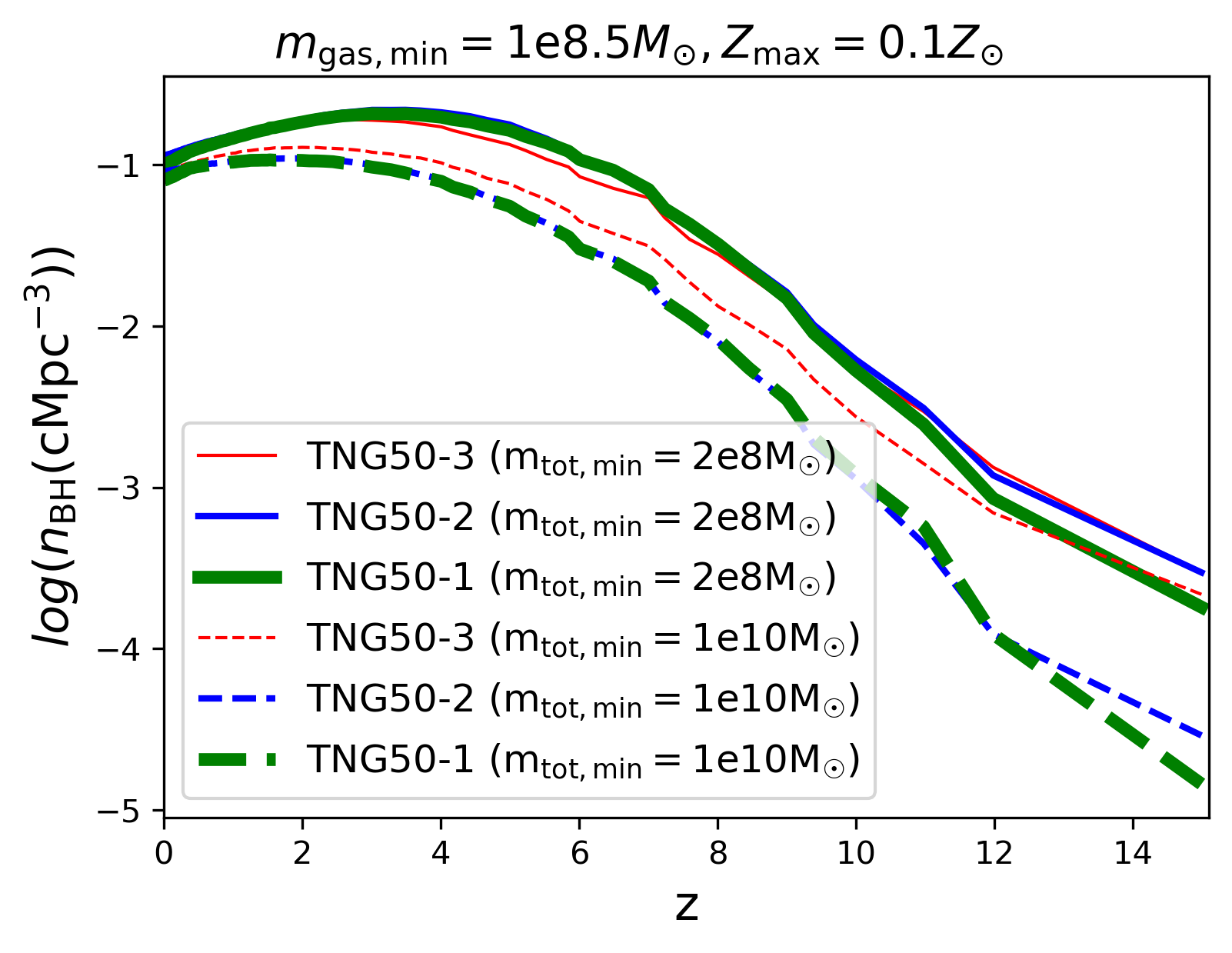}
\includegraphics[width=\columnwidth]{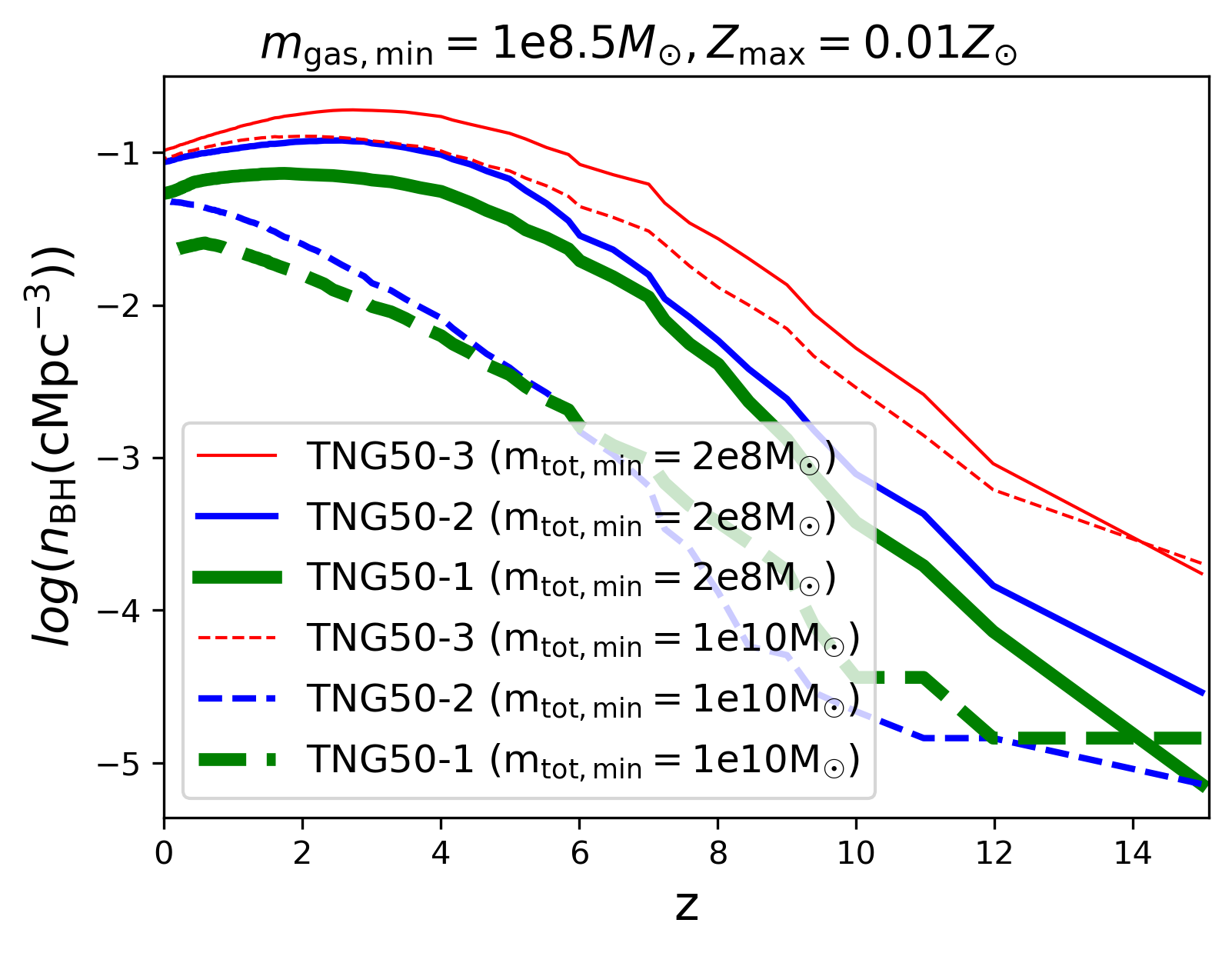}
\caption{ Resolution convergence is quantified through BH number density evolution. The results are shown for halos with $m_{\rm gas,min}=10^{8.5}$ \msun\ and either  $m_{\rm tot,min}=2\times10^8$\msun\ (in solid lines) or $m_{\rm tot,min}=10^{10}$ \msun\ (in dashed lines) with $\zgas/$\zsun~$=0.1$ and $\zgas/$\zsun~$=0.01$ host metallicity thresholds (top and bottom panels, respectively). Models with $\zgas/$\zsun~$=0.1$ in the top panel agree between resolutions to within $3-12$\% and to within $0.6-13$\% for the lower total mass cut. The lower-metallicity model (bottom panel) results converge more slowly with increasing resolution. The average gas-phase metallicity in halos at low resolutions is calculated from fewer gas cells. This makes it difficult to resolve the earliest enrichment of halos from their primordial metallicities. Therefore, we do not include metallicity thresholds below $Z_{\rm max} = 0.01$ \zsun\ in our seeding models.
\label{fig:res_convergence}}
\end{figure*}

In Figure~\ref{fig:res_convergence}, we assess the resolution convergence of our results by comparing the BH number density predictions of some of our semi-analytic seed models when applied to three TNG-50 volumes with different resolutions. We consider models with $m_{\rm gas,min}=10^{8.5}$ \msun\ and either  $m_{\rm tot,min}=2\times10^8$\msun\ or $m_{\rm tot,min}=10^{10}$ ( and either $\zgas/$\zsun~$=0.1$ or $\zgas/$\zsun~$=0.01$ shown in the top and bottom panels, respectively). The former has excellent resolution convergence (with results to within $\sim0.6-13$\% between resolutions for each mass cut), while the lower-metallicity model converges more slowly with increasing resolution. This occurs because at low resolutions, the average gas-phase metallicity in halos is calculated using a smaller number of gas cells. This limits our ability to resolve the earliest enrichment of halos from their primordial metallicities, and this fact informed our decision not to include metallicity thresholds below $Z_{\rm max} = 0.01$ \zsun\ in our seeding models. 

We also examine resolution convergence for the subset of BHs in halos with large total mass ($m_{\rm tot,min}=10^{10}$ \msun), dashed lines in Figure~\ref{fig:res_convergence}. The convergence is still excellent between TNG50-1 and TNG50-2 for the \texttt{mgas8.5\_Z0.1} model, while the lowest-resolution TNG50-3 simulation somewhat overpredicts number densities at high redshifts. In the \texttt{mgas8.5\_Z0.01} model, the subset of high-mass halos has similar resolution convergence as the full model results, except for TNG50-3, where BH number densities are vastly overpredicted. This indicates that high-mass halos are more susceptible to resolution effects when considering a maximum metallicity threshold, because in low-resolution simulations, enrichment of a few cells might quickly drive the average metallicity of small halos above $Z_{\rm max}$. In contrast, large halos with a few enriched cells are more likely to maintain reservoirs of pristine gas long enough for the BH seeding criteria to be met. Nonetheless, we still find reasonable agreement for the two highest-resolution simulations. In practice, this issue is largely moot in our study; as Figures~\ref{fig:number_densities.png} - \ref{fig:BHMFs.png} show, our low-$Z$, high-$m_{\rm tot}$ models are the ones that fail to produce a realistic BH population at any redshift.

\end{document}